\documentclass[usenatbib]{mnras}
\usepackage{newtxtext,newtxmath}

\usepackage[T1]{fontenc}

\DeclareRobustCommand{\VAN}[3]{#2}
\let\VANthebibliography\thebibliography
\def\thebibliography{\DeclareRobustCommand{\VAN}[3]{##3}\VANthebibliography}

\usepackage{graphicx}	
\usepackage{amsmath}	
\usepackage{caption}
\usepackage{subcaption}
\usepackage{booktabs}
\usepackage{makecell}
\usepackage{array}
\usepackage{soul}
\usepackage{rotating} 
\usepackage{pdflscape}
\usepackage{bm}
\usepackage{soul}
\usepackage{xcolor}

\newcommand{\boldifsmall}[1]{%
    \ifdim #1pt < 0.05pt
        \mathbf{#1}%
    \else
        #1%
    \fi
}

\newcommand{\num}[2]{#1\times 10^{#2}}
\newcommand{\bnum}[2]{\boldsymbol{#1\times 10^{#2}}} 
\newcommand{\ubnum}[2]{\underline{\boldsymbol{#1\times 10^{#2}}}} 

\renewcommand\arraystretch{1.6} 

\title[Colours of tidal features]{Tidal features as tracers of galaxy merger histories: the colours of tidal features in HSC-SSP}

\author[A. Desmons et al.]{
A. Desmons,$^{1}$\thanks{E-mail: a.desmons@unsw.edu.au}
S. Brough,$^{1}$
L. Canepa,$^{1}$
A. Khalid$^{1}$
\\
$^{1}$School of Physics, University of New South Wales, NSW 2052, Australia\\
}

\date{Accepted 2026 January 23. Received 2026 January 13; in original form 2025 November 09.}

\pubyear{2026}

\begin{document}
\label{firstpage}
\pagerange{\pageref{firstpage}--\pageref{lastpage}}
\maketitle

\begin{abstract}
We measure the radial $g-i$ colour profiles of $\sim$32,000 galaxies drawn from the Hyper Suprime-Cam Subaru Strategic Program optical imaging survey, including 1415 exhibiting tidal features. We compare the colour profiles of galaxies with and without tidal features to extract information about the properties of the mergers that created these features. We find negative colour gradients for both galaxies with and without tidal features and find that tidal feature-hosting red sequence galaxies have redder outskirts than their non-tidal feature hosting counterparts, consistent with the outskirts of these galaxies being dominated by stars accreted from gas-poor minor mergers. We find decreasing mass ratios of tidal features-to-host galaxy with increasing galaxy stellar mass, suggesting that less massive galaxies undergo mergers with companions closer in mass than more massive galaxies. Galaxies exhibiting streams have bluer outskirts than those hosting shells, and shells around red sequence galaxies tend to be more massive and have higher mass ratios to their hosts than streams, consistent with streams being formed from mergers with satellites less massive than those responsible for shells. The agreement between our findings and those of other observational and simulation-based works confirms the validity of our methodology and highlights the value of tidal features colours as a probe into the process through which galaxies evolve.  
\end{abstract}




\section{Introduction}
\label{sec:intro}
In the hierarchical structure formation model of the Universe, mergers play a significant role in the growth and evolution of galaxies, especially since $z\lesssim2$ (e.g. \citealt{Oser2010GalFormationPhases}). In particular, the growth of the most massive galaxies is dominated by the accretion of stellar material through mergers with lower mass galaxies (e.g. \citealt{Lacey1994NBodyMergeRate, Cole2000Hierarchical, Oser2010GalFormationPhases, Oser2012ETG_accretion, Robotham2014GAMAClosePair, Martin2018MergeMorphTransform, Remus2022MagneticumAccMass}). Ongoing and past mergers leave evidence of their interactions in the form of diffuse, non-uniform regions of stars in the outskirts of galaxies, known as tidal features. These features are composed of stellar material pulled out from the host or satellite galaxy during interactions (e.g. \citealt{Toomre1972BridgeTails}) and are expected to remain observable for $\sim3$ Gyr \citep{Mancillas2019ETGS_merger_hist, Huang2022HSCTidalFeatETG}. Tidal features contain valuable information about the merging history of a galaxy, and study of their colour and morphology can provide insights into the galaxy evolution process, including information about the mass ratio and angular momentum involved in mergers.

Common tidal feature morphologies include shells, streams, and tails. Shells are concentric arcs which curve around a host galaxy and are thought to originate primarily from radial mergers and mergers with low angular momentum (e.g. \citealt{Malin1983EllipGalShells, Quinn1984ShellsEllipGals, Hendel2015TidalDebOrbit, Pop2018ShellsIllustris, Valenzuela2024_TF_kins_form_hist}). They can originate from both minor and major mergers (e.g. \citealt{Pop2018ShellsIllustris, KadoFong2018HSCTidalFeat}) and are more commonly found around more massive early-type galaxies galaxies (e.g. \citealt{Atkinson2013CFHTLSTidal, Pop2018ShellsIllustris, KadoFong2018HSCTidalFeat, Desmons2023GAMA, Khalid2024TF_Sims, Rutherford2024MNRAS_SAMI_TF, Yoon2024A_ETG_rot, Desmons2025HSC_TF}). Streams are narrow filaments formed from infalling lower mass satellite galaxies on more circular orbits and are associated with minor mergers (e.g. \citealt{Hendel2015TidalDebOrbit, Mancillas2019ETGS_merger_hist, Sola2022TailsvsStreams}). Tails are elongated structures which, unlike streams, are composed of stellar material pulled out from a host galaxy and tend to be thicker and brighter than streams (e.g. \citealt{Sola2022TailsvsStreams, Pippert2025TailStreamModel}). They are associated with major merger origins and have shorter survival times than shells or streams \citep{Mancillas2019ETGS_merger_hist}.

In addition to the information about merger dynamics that can be gleaned from the morphology of tidal features, their colours can also reveal information about the mergers than created them. Using the relationship between galaxy colour and stellar mass (e.g. \citealt{Taylor2011GAMAMassEst}) the colour of tidal features can be used to estimate the stellar mass they contain (e.g. \citealt{Sola2025Strrings_streams}) and therefore the mass ratio of the mergers that created them (e.g. \citealt{KadoFong2018HSCTidalFeat}). By comparing the colours of shells to that of their host galaxies, \citet{KadoFong2018HSCTidalFeat} were able to determine that the majority of shells in their sample were consistent with minor merger origins, having $g-i$ colours bluer than their hosts, while $\sim15$ per cent of shells with colours similar to their hosts were consistent with major merger origins. Similarly, \citet{Sola2025Strrings_streams} used the $g-r$ colours of 35 streams around both early and late type galaxies, with stellar masses $\sim7\times10^9-3\times10^{11}\mathrm{M_{\odot}}$, to calculate the stellar masses of the streams and estimate the mass ratio of the mergers responsible for them, determining that the streams were consistent with minor merger origins.

In this work, we follow the methodology of \citet{Khalid2025TFColours}, who used mock images generated using simulated data to compare the $g-i$ colour profiles of galaxies with and without tidal features. We use the sample of $\sim$34,000 Hyper Suprime-Cam Subaru Strategic Program (HSC-SSP; \citealt{Aihara2018HSCSurveyDesign}) galaxies classified by \citet{Desmons2025HSC_TF} to study the colours of tidal features. By comparing the colour profiles of galaxies with and without tidal features, we can isolate the contribution of the tidal features to the profiles, and hence draw conclusions about the properties of the mergers responsible for their creation. Section~\ref{sec:Methods} details our sample selection and methods, including details of the tidal feature catalogue and how we measure galaxy sizes and colour profiles. Our results are presented in Section~\ref{sec:results}, where we compare the colour profiles of galaxies with and without tidal features, analyse the colour differences between individual classes of tidal features, and estimate the stellar mass contained in tidal features. In Section~\ref{sec:disc} we discuss our results, presenting physical explanations for our observations, and comparisons with other observational and simulation-based works. Finally, Section~\ref{sec:conc} presents our summary and conclusions. Throughout this paper, we assume a flat $\Lambda$CDM cosmology with $h=0.7$, $H_0=100~h~\rm{km}~\rm{s}^{-1}~\rm{Mpc}^{-1}$, $\Omega_m=0.3$,and $\Omega_{\Lambda}=0.7$.

\section{Methods}
\label{sec:Methods}

\subsection{Data sources and tidal feature catalogue}
\label{sec:data}
In this work, we use the pre-existing dataset assembled by \citet{Desmons2025HSC_TF}, composed of 34,331 galaxies sourced from the second Public Data Release of the HSC-SSP (PDR2; \citealt{Aihara2019HSCSecondData}) as a parent sample. The HSC-SSP survey \citep{Aihara2018HSCSurveyDesign} is a three-layered, \textit{grizy}-band imaging survey carried out with the Hyper Suprime-Cam on the 8.2m Subaru Telescope located in Hawaii, and the PDR2 has a median $\it{i}$-band seeing of 0.6 arcsec and a spatial resolution of 0.168 arcsec per pixel. The galaxies in the \citet{Desmons2025HSC_TF} dataset were sourced from the Deep/Ultradeep (D/UD) fields of the HSC-SSP, which span an area of $27$ and $3.5$~deg$^{2}$, respectively, and reach $\mu_{r}\sim$ 29.82~mag arcsec$^{-2}$ ($3\sigma,~10''\times10''$; \citealt{MartinezLombilla2023GAMAIntragroupLight}). This dataset was limited to galaxies with $i$-band magnitudes $15\leq~i~\leq~20$~mag, photometric redshifts $z~\leq~0.4$, and stellar masses $\mathrm{log}_{10}(M_{\star}/\mathrm{M_{\odot}})\geq9.5$. The photometric redshifts and stellar masses in \citet{Desmons2025HSC_TF} are the same used in this work, and were obtained from the \textsc{mizuki} catalogue available from the HSC-SSP database. These properties are inferred from the \textsc{mizuki} \citep{Tanaka2015PhotoZBayesionPrior} template-fitting code, which generates templates using stellar population synthesis models. Galaxy images in the \citet{Desmons2025HSC_TF} dataset were also required to have at least three exposures in each band ($g,~r,~i,~z,~y$), and objects flagged for cmodel fit failures, bad photometry, or flagged as being affected by bright sources were removed from the dataset.

To identify and classify tidal features in their dataset, \citet{Desmons2025HSC_TF} used a combination of machine learning and visual classification. The machine learning model was used to partially automate the detection of tidal features and reduce the visual classification workload. The model used was the pre-trained model designed by \citet{Desmons2024}, consisting of a self-supervised encoder which output lower-dimensional encodings of the input images, and a linear classifier which assigned scores between 0 and 1 to these encodings based on the probability of tidal features being present. \citet{Desmons2025HSC_TF} applied this model to $128\times128$ pixel images of the 34,331 galaxies in the dataset (resized from $256\times256$ pixel images using interpolation of pixel fluxes) to obtain a classifier score for each galaxy. \citet{Desmons2025HSC_TF} then visually classified the 10,000 galaxies assigned the highest classifier scores to ensure that the tidal feature sample was pure and not contaminated by galaxies without tidal features. The visual classification scheme used by \citet{Desmons2025HSC_TF} was similar to those used in \citet{Bilek2020MATLASTidalFeat}, \citet{Martin2022TidalFeatMockIm}, \citet{Desmons2023GAMA}, and \citet{Khalid2024TF_Sims}, and included the following tidal feature categories:
\begin{itemize}
    \item Shells: Concentric radial arcs or ring-like structures around a galaxy.
    \item Tidal tails: Prominent, elongated structures pulled out from the host galaxy. These usually have similar colours to that of the host galaxy.
    \item Stellar streams: Narrow filaments orbiting a host galaxy. These differ from tails in that they do not consist of material unbound from the host galaxy but instead consist of stars stripped from a smaller companion galaxy.
    \item Plumes or asymmetric stellar haloes: Diffuse features in the outskirts of the host galaxy, lacking well-defined structure like stellar streams or tails, or galaxies where the structure of the stellar halo is clearly asymmetric.
\end{itemize}

From their visual classification, \citet{Desmons2025HSC_TF} identified 1646 ($4.8$ per cent) galaxies with tidal features. These are the galaxies that make up our tidal feature sample for this work, with the remaining 32,685 galaxies making up our non-tidal feature sample. Since the non-tidal feature sample is primarily composed of galaxies not visually classified, it will likely contain some contamination from galaxies with tidal feature that could impact our results. Although the level of contamination is low (upper limit $8.3$ per cent; \citealt{Desmons2025HSC_TF}) we ensure that our results remain qualitatively unchanged when replacing the non-tidal feature sample with only the 8354 visually-confirmed non-tidal galaxies. In their work, \citet{Desmons2025HSC_TF} used images downloaded in all five bands $(g,r,i,z,y)$ available from the HSC-SSP survey. However, here we focus on the $g-$ and $i-$band images. We access the HSC-SSP galaxy images using the \textsc{Unagi} \textsc{Python} tool \citep{Huang2019Unagi} downloading 256~$\times$~256 pixel ( 42~$\times$~42 arcsec) images in each of these bands, centred around each galaxy.

Our work includes an analysis of whether rates of galaxies belonging to groups or clusters can account for some of the observed trends. When conducting such analyses we use the subsample from \citet{Desmons2025HSC_TF} which identifies galaxies in the parent sample of 34,331 galaxies that belong to groups or clusters. This subsample was assembled using the optically-selected CAMIRA cluster catalogue \citep{Oguri2018CAMIRA} and the spectroscopically-selected GAMA Galaxy Group Catalogue \citep{Robotham2011GAMAGroups}. It contains 492 GAMA members belonging to 89 distinct GAMA groups, spanning halo masses $12.6\leq\mathrm{log}_{10}(M_{200}/\mathrm{M}_{\odot})\leq14.6$, and 1871 CAMIRA members belonging to 83 distinct CAMIRA clusters, spanning halo masses $13.8\leq\mathrm{log}_{10}(M_{200}/\mathrm{M}_{\odot})\leq14.9$. 

\subsection{Galaxy size and colour measurements}
\label{sec:measurements}

\subsubsection{Masking and background subtraction}
\label{sec:masking}
In order to model the radial profiles of stars and to make accurate measurement of the size and colours of the galaxies in our sample, we must determine which light is associated with sources of interest and mask any light from other sources. Due to the size of our sample, our masking process must be fully automated and any parameters associated with the masking process must work for the entire sample. To detect and deblend sources in our sample we use the \texttt{detect\_sources} and \texttt{deblend\_sources} functions from the \textsc{photutils} \citep{Bradley2023zndoPhotutils} segmentation and watershed deblending tools. The data from HSC-SSP are already coadded, and we use these functions on $gi$ stacked images of the stars or galaxies as this stacking enhances the brightness of the sources and ensures that the masks are identical in all bands. The \texttt{detect\_sources} function detects groups of pixels composed of at least $N$ connected pixels each with a flux above a specified threshold, where pixels are considered connected if they are joined along the edges or corners. The segmentation map of detections output by this function is given as input to the \texttt{deblend\_sources} function which uses multi-thresholding and watershed segmentation to separate overlapping sources. 

As our data is sourced from the same survey and layer, we use the method and function parameters used by \citet{MartinezLombilla2023GAMAIntragroupLight}, except for the visual inspection component. Their method uses a cold~+~hot masking approach (e.g. \citealt{Rix2004HotColdMask, Montes2018HotColdMask, Montes2021A85_ICL}), where the cold mask captures bright, extended objects (e.g. galaxies and bright stars), and the hot mask captures faint, small objects (e.g. background galaxies). To improve source detection, images are convolved with a Gaussian filter (\texttt{Gaussian2DKernel} from \textsc{astropy}) with $\sigma=5$ pixels before the cold or hot masking process. Additionally, the hot masking process is run on an unsharp masked version of the original image. This is done by subtracting the convolved version of the image from the original image. The \texttt{detect\_sources} and \texttt{deblend\_sources} functions are applied with \texttt{npixels}~=~40 and \texttt{npixels}~=~7 for the cold and hot masks, respectively. We set the \texttt{threshold} parameter to  $1.1\sigma_{\rm{coadd}}$ and $1.1\sigma_{\rm{unsharp}}$ for the cold and hot masking, where $\sigma_{\rm{coadd}}$ and $\sigma_{\rm{unsharp}}$ are the background noise level of the coadded $gi$ and unsharp images, respectively, given by the \texttt{detect\_threshold} \textsc{photutils} function. This function uses sigma-clipped statistics to calculate a scalar background and noise estimate. To ensure that the faint outskirts of sources are completely covered by the masks, we expand the cold and hot masks by convolving them with Gaussian filters with $\sigma=2$ and $\sigma=1$, respectively.

\begin{figure}
    \centering
    \includegraphics[width=0.7\columnwidth]{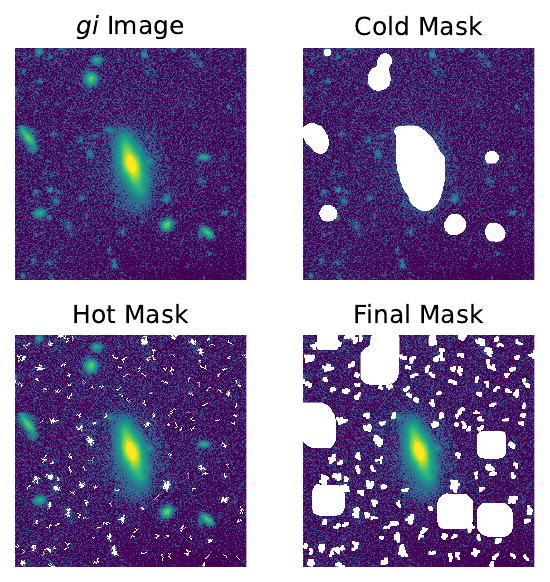}
    \caption{Example of the masks generated using our cold and hot mask approach on a galaxy from our sample. Top left: the $gi$ coadded image of the galaxy. Top right: the cold mask before being expanded. Bottom left: the hot mask before being expanded. Bottom right: the final mask with the central galaxy unmasked, after expanding the cold and hot masks by convolving them with Gaussian kernels.}
    \label{fig:masking}
\end{figure}

\begin{figure*}
    \centering
    \includegraphics[width=0.95\textwidth]{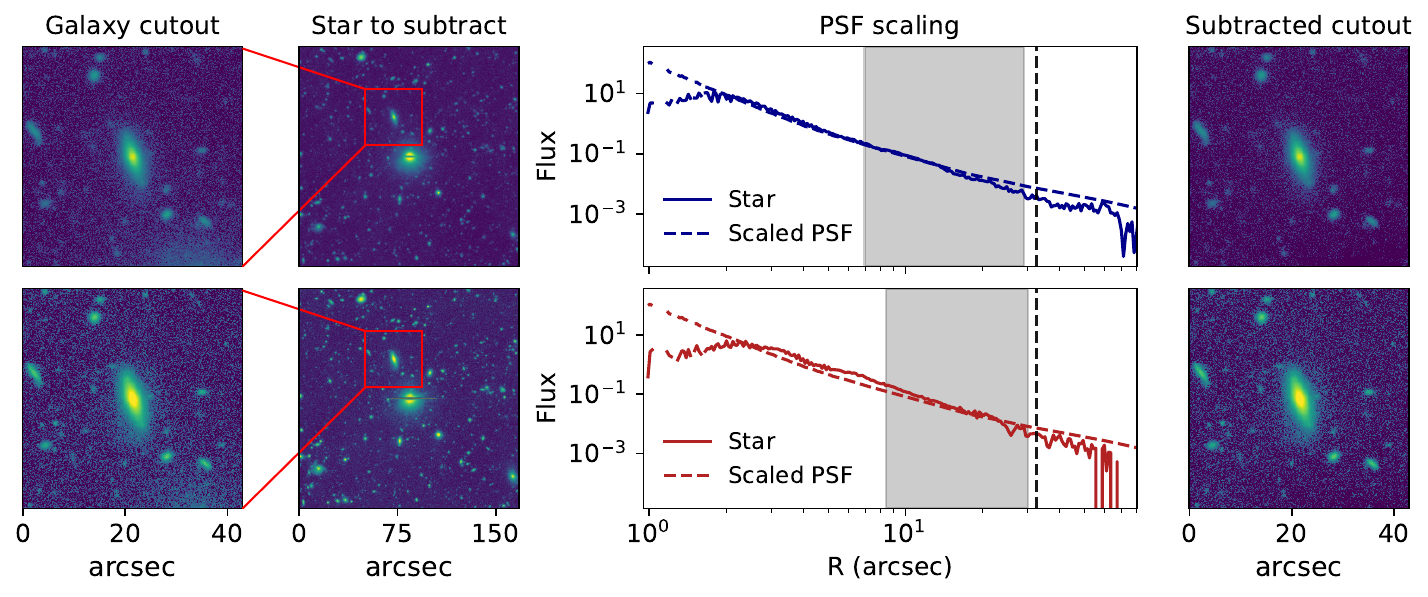}
    \caption{Example of our scattered light subtraction process for a galaxy with a nearby star. The top and bottom rows show the same process in $g-$ and $i-$band, respectively. The left panels show the $256\times256$ pixel cutout of the galaxy, and the middle-left panels show a 1000 pixel$^2$ cutout (cropped from the original 3400 pixel$^2$ cutout for visibility) centred on the star to be subtracted. The middle-right panels show the profile of the final scaled PSF (dashed lines) compared to the profile of the star (solid lines). The grey shaded region in these panels show the radial range used when calculating the PSF scale factor, and the dashed vertical lines show the distance of the galaxy relative to the star. The right panels show the final $256\times256$ pixel cutout once the scattered light from the star has been subtracted.}
    \label{fig:star_sub}
\end{figure*}

Certain parts of our analysis require specific sources such as stars or galaxies to be unmasked from an image. Since each source in the cold mask segmentation map is assigned a unique value, we do this by identifying the segmentation map value at the location of our source and removing the mask from all pixels with this value. Occasionally, a star or galaxy is deblended into multiple sources. For stars, this usually happens due to saturation spikes, an example of which can be seen in the second panels of Fig.~\ref{fig:star_sub}, where a spike runs horizontally through the centre of the star. To address this issue, we add a dither factor $N_{\rm{dither}}$ to our unmasking process such that not only the segmentation map value at the location of our source at ($x,y$) is unmasked, but also the segmentation map value around the source at ($x~\pm~N_{\rm{dither}},y$) and ($x,y~\pm~N_{\rm{dither}}$). This factor is set to $N_{\rm{dither}}=10$ pixels when unmasking stars, and $N_{\rm{dither}}=2$ pixels when unmasking galaxies. Fig.~\ref{fig:masking} illustrates the masking process on a galaxy from our sample.

While the HSC-SSP data are already sky subtracted, the two-step global sky subtraction carried out by the pipeline is performed on spatial scales of $1024\times1024$ pixels ($\sim2.8'\times2.8'$) and $256\times256$ pixels ($43''\times43''$). To eliminate any residual background, or gradients, that may remain in our $256\times256$ pixel images and affect our size and colour measurements, we therefore also model and subtract a 2D background from these images (e.g. \citealt{Li2022MassiveGalOutskirt, MartinezLombilla2023GAMAIntragroupLight}). This is done using the \texttt{Background2D} function from the \textsc{photutils} package, which places a grid of boxes throughout an image and estimates the background using sigma-clipped statistics in each box. This low-resolution background map is then interpolated to obtain a final background map for the image. We apply the \texttt{Background2D} function on the masked versions of our images, using 16 $10\times10~\rm{arcsec}^{2}$ boxes. The \texttt{exclude\_percentile} variable in the \texttt{Background2D} function determines what percentage of pixels in a box can be masked before it is excluded from the background calculation. To increase the reliability of the measured background, we iterate over 3 values of \texttt{exclude\_percentile} [25, 50, 75] until we obtain a background map generated using at least 10 boxes. If no background measurement can be obtained using at least 10 boxes we discard the galaxy from our sample; this is the case for 1215 galaxies. These 1215 galaxies do not have significantly different stellar mass, $g-i$ colour, or effective semi-major radius distributions from the sample of galaxies we successfully model, and their removal does not introduce any bias into our final sample. However, these 1215 galaxies are more likely to belong to groups and clusters ($8.6_{-0.7}^{+0.9}$ per cent) than the successfully modelled galaxies ($6.8\pm0.1$ per cent). This is expected, as group and cluster environments will be more populated and hence require a greater portion of the image to be masked, reducing the number of boxes available to model the background. 

\subsubsection{Scattered light subtraction}
\label{sec:star_sub}
Our goal is to make accurate measurements of the brightness and colour of the outskirts of galaxies. However, the scattered light from bright foreground stars can add an artificial flux component to the actual amount of light in the galaxies. This can have a significant impact on the measured brightness and colour of galaxies, especially in the faint outskirts of galaxies (e.g. \citealt{Uson1991PSFcontam, Michard2002PSFcontam, Slater2009PSFcontam, Sandin2014PSFcontam, Sandin2015PSFcontam}). We therefore want to minimise the impact of scattered light from nearby stars in our images. Since light scatters differently in the $g-$ and $i-$bands, we need to model the stars affecting our galaxies in these bands independently, and subtract them from our images to get an accurate measurement of the $g-i$ colour of galaxies and their outskirts. We do this by following the method outlined in \cite{MartinezLombilla2023GAMAIntragroupLight}, using an extended point spread function (PSF) of the survey to model the scattered light from stars. 

We use the $g-$ and $i-$band extended PSFs constructed by \citet{MartinezLombilla2023GAMAIntragroupLight} using data from the wide layer of the HSC-SSP PDR2. These PSFs were built following the method outlined in \citet{Infante-Sainz2020PSFs}, using median-clipping star stacking and point-like sources in three different ranges of brightnesses to build the inner, intermediate, and outer part of the PSF. The PSF in each band extends to $\sim1600$ pixels, equivalent to 4.5~arcmin.

To obtain the locations of stars that may affect the galaxies in our sample we use the star catalogue constructed by \citet{Coupon2018HSCBrightStars}. This catalogue was assembled using data from the Gaia first data release \citep{Gaia_collab2016Gaia, Gaia_collab2016GaiaDR1} and the Tycho-2 star sample \citep{Hog2000Tycho2}, and contains the positions of all stars brighter than $G_{\rm{Gaia}}=18$ which overlap with the HSC-SSP footprint. We cross-match this star catalogue with our sample of 34,311 galaxies, keeping only the 31,549 stars which are within 4.5~arcmin (the extent of the PSF model) of any of these galaxies. We also apply a magnitude-dependent distance cut on the distance of stars from galaxies. We do this by identifying the distance from the centre of a star at which its surface brightness profile reaches depths fainter than the survey limit for a range of star magnitudes. Table~\ref{tab:star_dist_cut} shows these distances for each of the star magnitude bins we consider. This reduces our sample to 16,001 stars, affecting 27,470 galaxies. We download 3400 pixel$^2$ (9.5~arcmin) HSC-SSP images for 15,809 of these stars. The \citet{Coupon2018HSCBrightStars} star catalogue contains stars which are up to $1^{\circ}$ beyond the footprint of the HSC-SSP survey, therefore images of these stars could not be obtained. This is the case for 192 stars, affecting 506 galaxies in our sample. Although these stars could not be modelled and subtracted, we do not discard the 506 galaxies from our sample and instead verify that our conclusions remain unchanged if they are removed.

\begin{table}
\renewcommand\arraystretch{1.2}
\centering
\begin{tabular}{ccc}
\hline
$\mathrm{G}_{\mathrm{GAIA}}$ (mag)& 
$d_{\mathrm{max}}$ (arcsec) &
$N_{\mathrm{star}}$\\ \hline
$\mathrm{G}\leq11$ & 270 & 372 \\
$11<\mathrm{G}\leq12$ & 220 & 535 \\
$12<\mathrm{G}\leq13$ & 165 & 1069 \\
$13<\mathrm{G}\leq14$ & 145 & 1895 \\
$14<\mathrm{G}\leq15$ & 115 & 2317 \\
$15<\mathrm{G}\leq16$ & 90 & 3969 \\
$16<\mathrm{G}\leq17$ & 65 & 4261 \\
$\mathrm{G}>17$ & 30 & 1391 \\
\hline
\end{tabular}
\caption{Maximum projected distance ($d_{\mathrm{max}}$) stars of a given magnitude ($\mathrm{G}_{\mathrm{GAIA}}$) can be from galaxies, to be considered for subtraction from an image. $N_{\mathrm{star}}$ indicates the final number of stars to be subtracted in each magnitude range.}
\label{tab:star_dist_cut}
\end{table}

To scale the PSF model to each of these stars we follow the method outlined in \citet{Roman2020star_sub, Infante-Sainz2020PSFs, Montes2021A85_ICL} and \citet{MartinezLombilla2023GAMAIntragroupLight} with some variations to account for the wider range of star brightnesses in our sample, and the fact that our method must be fully automated. Given an image of a star in $g-$ or $i-$band, we begin by masking all sources in the image other than the star, following the masking procedure detailed in Section~\ref{sec:masking}. We then obtain radial profiles of the flux of the star and the PSF model using the \textsc{CircularAnnulus} task from the \textsc{photutils} package \citep{Bradley2023zndoPhotutils}, which applies logarithmically spaced circular apertures and calculates the average flux in each annulus after applying 3$\sigma$ clipping. We then calculate a scale factor for the star, which is the ratio between its radial profile and the profile of the corresponding PSF model. When calculating the scale factor, we use a limited radial region rather than using the full radial profile of the star. The inner limit on the radius ensures that we avoid regions oversaturated by the core of the star, and the outer limit ensures we only use regions of the profile before background noise begins to dominate. The radial range we use spans 0.01 times the maximum flux of the star to 2 times the background value of the image. This background value is calculated by taking the median flux of the image after masking all sources. 

To subtract the scattered light from stars affecting a given galaxy in our sample, we start by obtaining the scale factors in $g-$ and $i-$band for each of the stars affecting the galaxy. We then scale the PSF model for each star in each band using the corresponding scale factor, and extract the $256\times256$~pixel region of the scaled PSF where the galaxy image would be. We sum these regions for each star affecting the galaxy in $g-$ and $i-$band and subtract this from our $g-$ and $i-$band galaxy images, respectively. Fig.~\ref{fig:star_sub} illustrates this method, showing the radial profile of a PSF scaled to a star affecting a galaxy, and the galaxy image after subtracting the scattered light in $g-$ and $i-$bands.

Appendix \ref{sec:app_star_sub} provides an analysis of the effect of star subtraction on our galaxy size and colour measurements. We also provide a version of the colour profiles in Section~\ref{sec:colour_prof}, this time without star subtraction, and find that our results remain qualitatively unchanged.

\begin{figure}
    \centering
    \includegraphics[width=0.99\columnwidth]{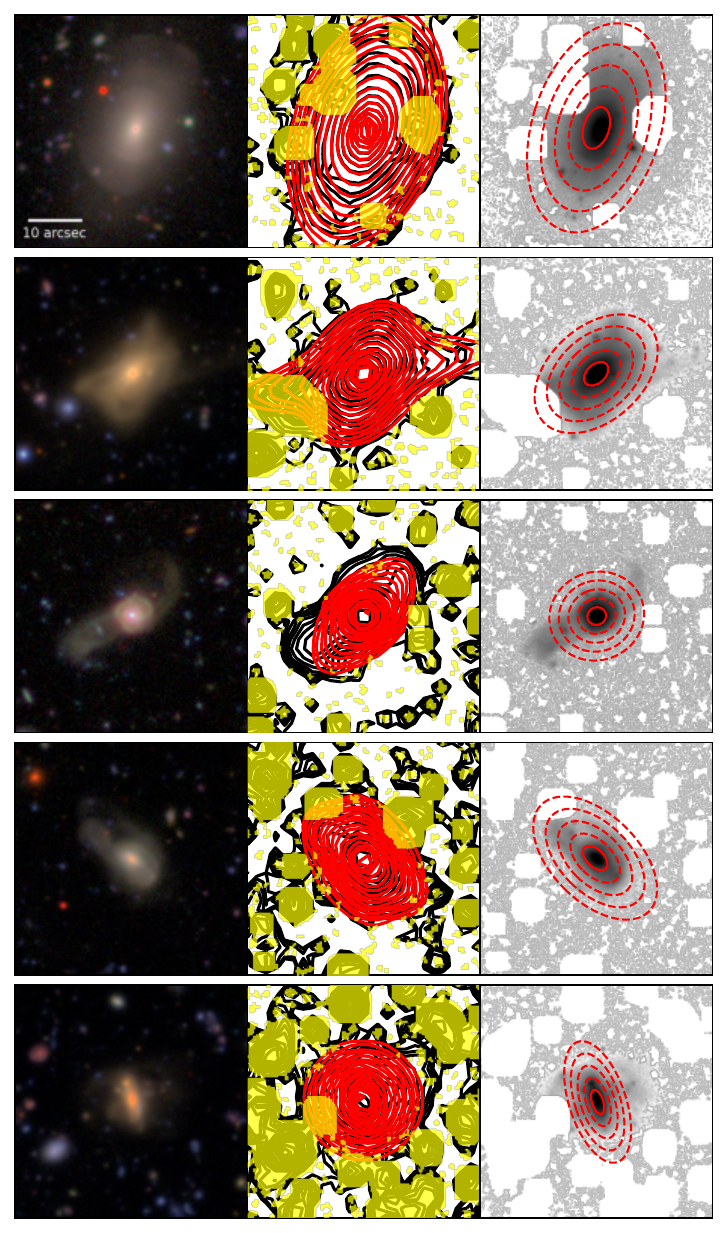}
    \caption{Illustration of the method used to measure galaxy size and colour. Left column: arcsinh stretched $gri$ false colour image showing examples of each type of tidal feature in the \citet{Desmons2025HSC_TF} classification scheme. From top to bottom: shell, stream, tail, plume, and double nucleus. Middle column: MGE fit (red contours) to the data (black contours), with the yellow showing masked regions. The fits are performed on $gi$ coadded images of the galaxies in the left column. Right column: masked grey-scale $gi$ coadded images and the apertures at  $1R_e^{\rm{maj}}$ (solid red) and  $2, 3,4,\mathrm{and}~5R_e^{\rm{maj}}$ (dashed red) used to measure the colour profiles of the galaxies. The apertures in this column are based on the average ellipticity and position angle of the galaxy, as given by \texttt{find\_galaxy}.}
    \label{fig:MGE_demo}
\end{figure}

\begin{figure*}
    \centering
    \includegraphics[width=0.95\textwidth]{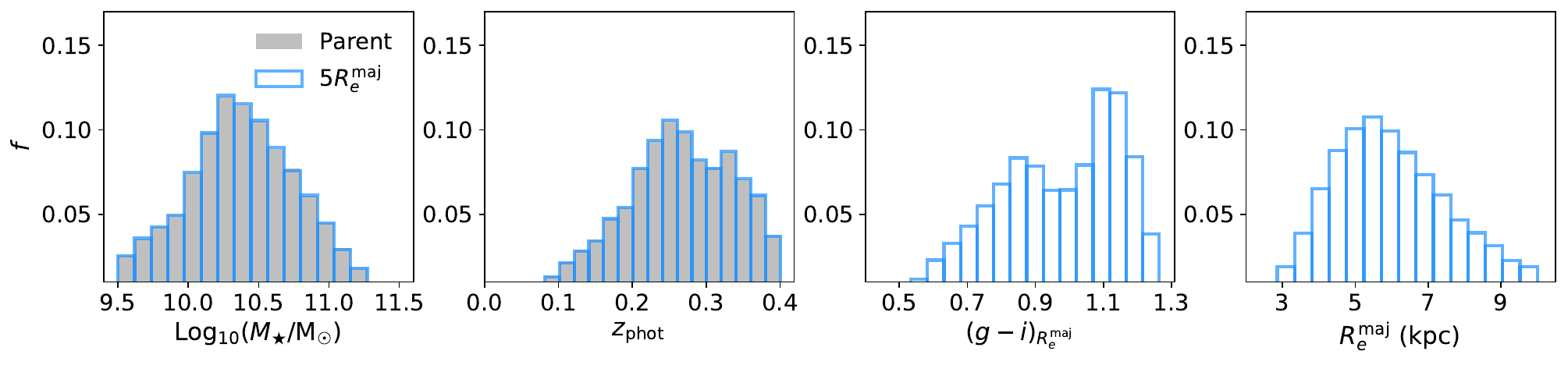}
    \caption{The first two panels show the distribution of stellar mass and photometric redshift for the parent sample of 34,331 galaxies (grey) and galaxies with reliable size and colour measurements out to $5R_e^{\rm{maj}}$ (blue). The last two panels show the distribution of measured $g-i$ colour within $1R_e^{\rm{maj}}$ and effective radius for galaxies with reliable size and colour measurements out to $5R_e^{\rm{maj}}$.}
    \label{fig:dist_hists}
\end{figure*}

\subsubsection{Size measurements}
\label{sec:size_measure}
We measure the sizes of galaxies in our sample by closely following the method detailed in \citet{Khalid2025TFColours} which involves using the Multi-Gaussian Expansion fitting algorithm \citep{Monnet1992MGE, Emsellem1994MGE} implemented in the \texttt{MGEfit} package \citep{Cappellari2002MGE}. The size measurements are performed on the $gi$ coadded $256\times256$ pixel images, with the scattered light from nearby stars subtracted as in Section~\ref{sec:star_sub}, the background subtracted, and all sources apart from the central galaxy masked as detailed in Section~\ref{sec:masking}. To measure the average ellipticity and position angle of the galaxies we use the \texttt{find\_galaxy} function which identifies the brightest region of connected pixels lying above a specified brightness threshold. It then calculates the galaxy centre, position angle, and ellipticity of a galaxy using the first and second-order moments of the intensity distribution of this region. The brightness threshold used by \texttt{find\_galaxy} is determined by the \texttt{fraction} variable, which indicates the fraction of the brightest pixels in the image that should be used for the search. Given the range of sizes and brightnesses of galaxies in our sample, a single value of \texttt{fraction} does not work for the entire sample. We find that lower values of \texttt{fraction} are useful for separating interacting galaxies or galaxies with close neighbours, while higher values of \texttt{fraction} are better suited for fainter galaxies. We therefore run \texttt{find\_galaxy} on each galaxy, iterating over 5 values of \texttt{fraction} [0.015, 0.025, 0.05, 0.1, 0.25], until a suitable solution is found. Since our images are centred on the galaxies in our sample, we consider a solution suitable when the centre of the galaxy found by \texttt{find\_galaxy} is within 10 pixels of the image centre. If the algorithm fails to find a suitable solution after iterating over the  different values of \texttt{fraction} we discard the galaxy from our sample. This is the case for 31 galaxies, which upon visual inspection are either significantly overlapping with other galaxies nearby, or are impacted by imaging artefacts.  

We then model each galaxy using multiple Gaussian components using unregularized fits, which produce more realistic galaxy shapes (e.g. \citealt{D'Eugenio2021MGEUnreg}), as opposed to regularized fits, which are biased toward more circular shapes better suited for dynamical modelling (e.g. \citealt{Scott2009MGERegularise}). Following the procedure in \citet{Khalid2025TFColours}, we also allow the position angles of models to vary with radius, and to produce more accurate models of the galaxies, we also convolve the fits with circular MGE models of the extended $g-$band HSC-SSP PSF. Based on the MGE model and the average ellipticity and position angle from \texttt{find\_galaxy}, we can measure the effective semi-major radius $R_e^{\rm{maj}}$ of each galaxy, defined as half the semi-major axis of the elliptical aperture enclosing half of the galaxy's light. Some of the galaxies in our sample with very small extents on the sky could not be modelled due to the minor axis of the fit being smaller than the minimum dispersion of the Gaussians used in the PSF fitting ($0.2$~arcsec). This is the case for 452 galaxies, and since no reliable model can be obtained for these galaxies, they are discarded from the sample. Fig~\ref{fig:MGE_demo} shows examples of fits obtained for galaxies in our sample.

When measuring the colour of galaxy outskirts (see Section~\ref{sec:col_measure}), we consider elliptical apertures out to $5R_e^{\rm{maj}}$ around each galaxy, meaning that the elliptical apertures at $5R_e^{\rm{maj}}$ for galaxies with $R_e^{\rm{maj}}>25$~pixels will extend beyond the bounds of our $256\times256$ pixel images. This is the case for 740 galaxies in our sample. While many of these galaxies do appear large in the images, others were incorrectly modelled due to being very compact objects contaminated by light from neighbouring galaxies, making them impossible to model. We therefore visually inspect these 740 galaxies, discarding 275 galaxies for which reliable modelling is unable to be achieved. For the remaining 465 galaxies with measured $R_e^{\rm{maj}}>25$~pixels, we download larger versions of the images ($800\times800$~pixels) and repeat the size measurement and MGE modelling. We also discard 1 galaxy whose effective radius is measured to be smaller than half the FWHM of the $g-$band PSF (4.5 pixels) as this measurement is likely to be unreliable.

\subsubsection{Colour measurements}
\label{sec:col_measure}
The intrinsic $g-i$ colours of our galaxies are derived from the light enclosed within an elliptical aperture at $1R_e^{\rm{maj}}$, with the ellipticity and position angle measured by \texttt{find\_galaxy}. When calculating $g-i$ colour profiles we use this same ellipticity and position angle, and measure the light enclosed in apertures of semi-major axis $0-1R_e^{\rm{maj}}$, $1-2R_e^{\rm{maj}}$, $2-3R_e^{\rm{maj}}$, $3-4R_e^{\rm{maj}}$, and $4-5R_e^{\rm{maj}}$. We convert our $g-i$ colours to rest-frame measurements by applying k-corrections to each band using the methods outlined in \citet{Chilingarian2010MNRASkcorrect1} and \citet{Chilingarian2012MNRASkcorrect2}. 

\subsection{Sample selection}
\label{sec:sample}
After applying our star subtraction, masking, background subtraction, and MGE modelling procedures to the 34,331 galaxies in our parent sample, we manage to successfully model 32,357 galaxies. To ensure that our magnitude and colour measurements are performed on apertures with sufficient portion of the galaxy unmasked we discard 96 galaxies which are more than 50 per cent masked within the $1R_e^{\rm{maj}}$ aperture, reducing the sample to 32,261 galaxies, 1486 ($4.6$ per cent) of which have visually classified tidal features. When analysing the colour profiles of galaxies using apertures out to $5R_e^{\rm{maj}}$, we also discard galaxies that are more than 75 per cent masked in any of the 4 apertures from  $1-5R_e^{\rm{maj}}$. This removes a further 442 galaxies, resulting in a sample of 31,819 galaxies, 1415 ($4.4$ per cent) of which have tidal features. The 75 per cent masking threshold is chosen heuristically as we find no relationship between the $g-i$ colour measurement or scatter in $g-i$ colour in an aperture and the percentage of the aperture masked. We ensure that this choice of acceptable masking threshold does not impact our conclusions by confirming that our final results remain qualitatively unchanged when employing a lower threshold of 50 per cent masked in the $1-5R_e^{\rm{maj}}$ apertures, which discards 5187 galaxies. The first two panels of Fig.~\ref{fig:dist_hists} shows the distribution of stellar mass and photometric redshift for our parent sample of 34,331 galaxies and for the galaxies with reliable size and colour measurements out to $5R_e^{\rm{maj}}$, showing no significant differences between the samples. The last two panels of Fig.~\ref{fig:dist_hists} show the distribution of $g-i$ colour measured within $1R_e^{\rm{maj}}$ and radius for galaxies with reliable size and colour measurements out to $5R_e^{\rm{maj}}$. The median stellar mass, photometric redshift, $g-i$ colour, and effective radius for the $5R_e^{\rm{maj}}$ sample are $\mathrm{log}_{10}(M_{\star}/\mathrm{M_{\odot}})=10.4~\pm~0.4$, $z=0.26~\pm~0.08$, $g-i=1.0~\pm~0.2$, and $R_e^{\rm{maj}}=6~\pm~2$~kpc, where the uncertainties given are the median absolute deviation.

\begin{figure}
    \centering
    \includegraphics[width=0.99\columnwidth]{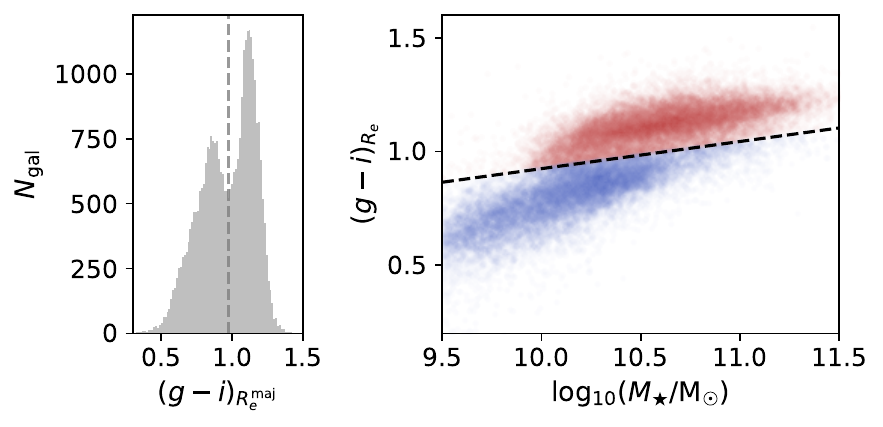}
    \caption{Left: distribution of $(g-i)_{R_{e}^{\mathrm{maj}}}$ colours in our sample, with the dashed line at $(g-i)_{R_{e}^{\mathrm{maj}}}=0.975$ showing our preliminary threshold between red and blue galaxies. Right: the red sequence and blue cloud populations in our sample, with the dashed line showing our final red vs. blue cut-off as defined by Equation~\ref{eq:col_cut}.}
    \label{fig:col_cut}
\end{figure}

To limit the impact that differences in colour profiles for red sequence and blue cloud galaxies (e.g. \citealt{Miller2023RedSeqBlueCloud}) may have on our colour profile analysis, we divide our galaxy sample into a red sequence and blue cloud population. We define this divide by first locating the local minimum in a $(g-i)_{R_{e}^{\mathrm{maj}}}$ histogram between the two peaks representing the red sequence and blue cloud, giving us a preliminary threshold between red and blue galaxies, illustrated in the left panel of Fig.~\ref{fig:col_cut}. We then fit the red sequence in colour-stellar mass space by linearly fitting the population of galaxies redder than this threshold, and calculate the median scatter of galaxies redder than the threshold from this red sequence fit. The division between the red sequence and blue cloud is then defined as 3 times this median scatter below the red sequence fit, and follows the relationship:
\begin{equation}
    g - i = 0.12 \times \mathrm{log_{10}}(M_{\star}/\mathrm{M_{\odot}}) - 0.27
    \label{eq:col_cut}
\end{equation}
This division is shown in Fig.~\ref{fig:col_cut}, along with the resulting red sequence and blue cloud populations. This results in a sample of 16,704 red sequence galaxies (986 with tidal features) and 15,115 blue cloud galaxies (429 with tidal features) which we use for our analysis comparing the colour profiles of galaxies with and without tidal features.

\begin{figure}
    \centering
    \includegraphics[width=0.9\columnwidth]{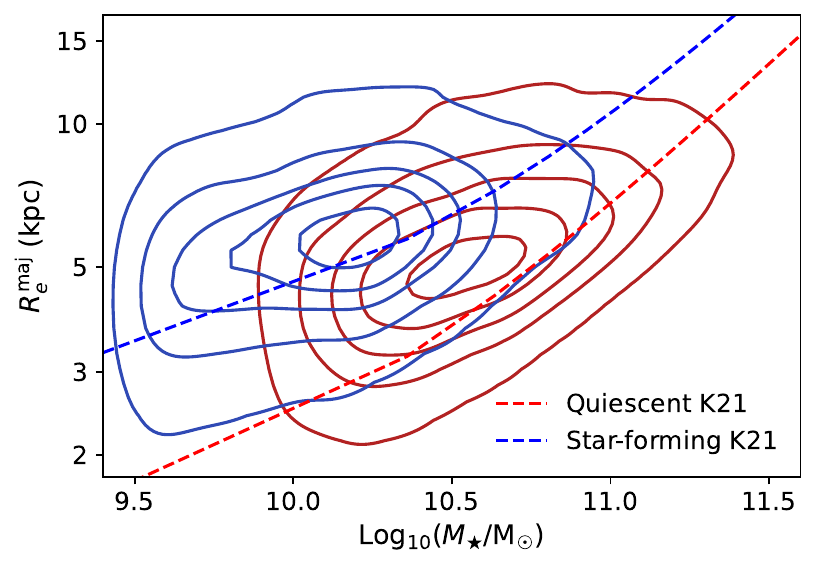}
    \caption{The size-stellar mass relation for galaxies which have reliable measurements out to $1R_e^{\rm{maj}}$ for the red sequence (red contours) and blue cloud (blue contours). The contours depict 10, 30, 50, 70, and 90 per cent of the data. The blue and red dashed lines show the empirical size-mass relations derived by \citet{Kawinwanichakij2021HSCMassSize} for star-forming and quiescent galaxies in HSC-SSP, respectively. Our size-mass relation is in agreement with those from \citet{Kawinwanichakij2021HSCMassSize}.}
    \label{fig:mass_rad_col}
\end{figure}

Fig.~\ref{fig:mass_rad_col} shows the size-stellar mass distribution for our samples of red sequence and blue cloud galaxies. We compare our size-stellar mass distribution to the size-stellar mass relations obtained for quiescent and star-forming galaxies in HSC-SSP at $z_{\rm{median}}=0.3$ \citep{Kawinwanichakij2021HSCMassSize} and find reasonable agreement, with any differences likely due to their specific separation of quiescent and star-forming galaxies rather than the more simple $g-i$ colour separation used here. The general agreement with the \citet{Kawinwanichakij2021HSCMassSize} relations supports the validity of our method to obtain galaxy sizes and colours.

The detectability of tidal features is dependent on the surface brightness, morphology, and orientation of the features and their host galaxies, potentially leading to incompleteness and consequent bias in our sample, correlated with host galaxy properties. For example, the fainter nature of streams (e.g. \citealt{Sola2022TailsvsStreams, Pippert2025TailStreamModel}), which are associated with older and lower mass ratio accretion events compared to tails (e.g. \citealt{Hendel2015TidalDebOrbit, Mancillas2019ETGS_merger_hist, Sola2022TailsvsStreams}), causes them to be harder to detect than tails. As a result, our sample of galaxies with detected tidal features may be biased toward more massive galaxies, more recent mergers, or mergers with higher mass ratios. Consequently, the sample should not be used to infer population-level trends such as the overall frequency of tidal features, the relative frequency of merger types, or the stellar mass dependence of recent and ongoing mergers. However, this should not affect the results of our analysis, as we focus on relative comparisons within our sample and match samples in stellar mass when possible.

\section{Results}
\label{sec:results}

In this Section, we present an analysis of the brightness and colour profiles of galaxies with tidal features, using the set of galaxies without tidal features as a control sample. We compare the colour profiles of galaxies with different categories of tidal features and estimate the stellar mass contained in those features. In addition to considering red sequence and blue cloud galaxies separately, we divide our sample into four stellar mass bins to  limit the impact of the colour-stellar mass relationship on our analysis.

\subsection{Galaxy brightness profiles}
\label{sec:brightness_prof}
\begin{figure}
    \centering
    \includegraphics[width=0.99\columnwidth]{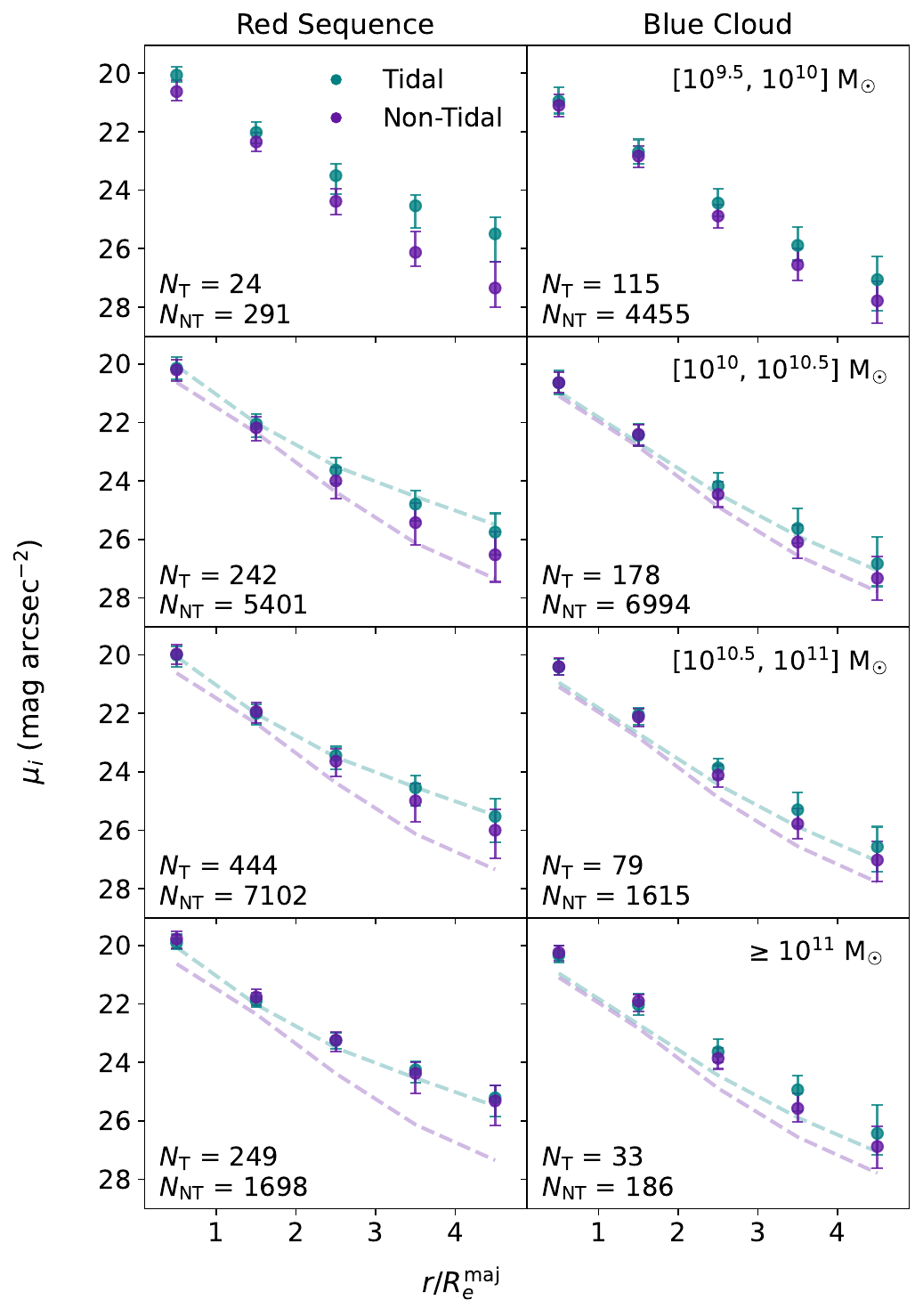}
    \caption{Distributions of radial brightness profiles for red sequence (left) and blue cloud (right) galaxies in four stellar mass bins, given in terms of $i-$band surface brightness ($\mu_i$) and the semi-major axis ($r/R_{e}^{\mathrm{maj}}$). The profiles are shown for galaxies with tidal features (teal) and without tidal features (purple), and are measured using elliptical apertures with integer spacing from 0 to 5~$R_{e}^{\mathrm{maj}}$. Each point shows the median $\mathrm{M}_i$ in that aperture, with the error bars depicting the $25^{\mathrm{th}}$ and $75^{\mathrm{th}}$ percentiles. The number of galaxies with and without tidal features used to generate the profiles are indicated in each panel by $N_{\mathrm{T}}$ and $N_{\mathrm{NT}}$, respectively. The dashed lines in each panel show the profiles for the lowest stellar mass bin to aid comparison between mass bins. Galaxies with tidal features tend to be brighter at radii beyond $2R_{e}^{\mathrm{maj}}$, suggesting that tidal features are contributing to the light in galaxy outskirts.}
    \label{fig:mu_i_r_re}
\end{figure}

In Fig.~\ref{fig:mu_i_r_re} we compare the brightness profiles of galaxies with and without tidal features using their $i-$band surface brightness. Blue cloud galaxies tend to be brighter at all radii with increasing stellar mass, which is not the case for red sequence galaxies. We find that, while tidal feature and non-tidal feature galaxies have comparable brightnesses within $1{R_{e}^{\mathrm{maj}}}$, the profiles tend to deviate at radii beyond $2{R_{e}^{\mathrm{maj}}}$, with tidal feature galaxies having brighter outskirts. While this is expected, as tidal features provide additional sources of light in the outskirts of galaxies, it confirms that our method is successfully allowing us to detect the additional light associated with tidal features. While not statistically significant on a per-bin basis, the systematic deviation in brightness between galaxies with and without tidal features tends to be strongest in lower mass galaxies, and decreases with increasing stellar mass for the red sequence, disappearing completely for the highest mass ($\mathrm{log}_{10}(M_{\star}/\mathrm{M_{\odot}})>11$) red sequence galaxies.

\subsection{Galaxy colour profiles}
\label{sec:colour_prof}
\begin{figure}
    \centering
    \includegraphics[width=0.99\columnwidth]{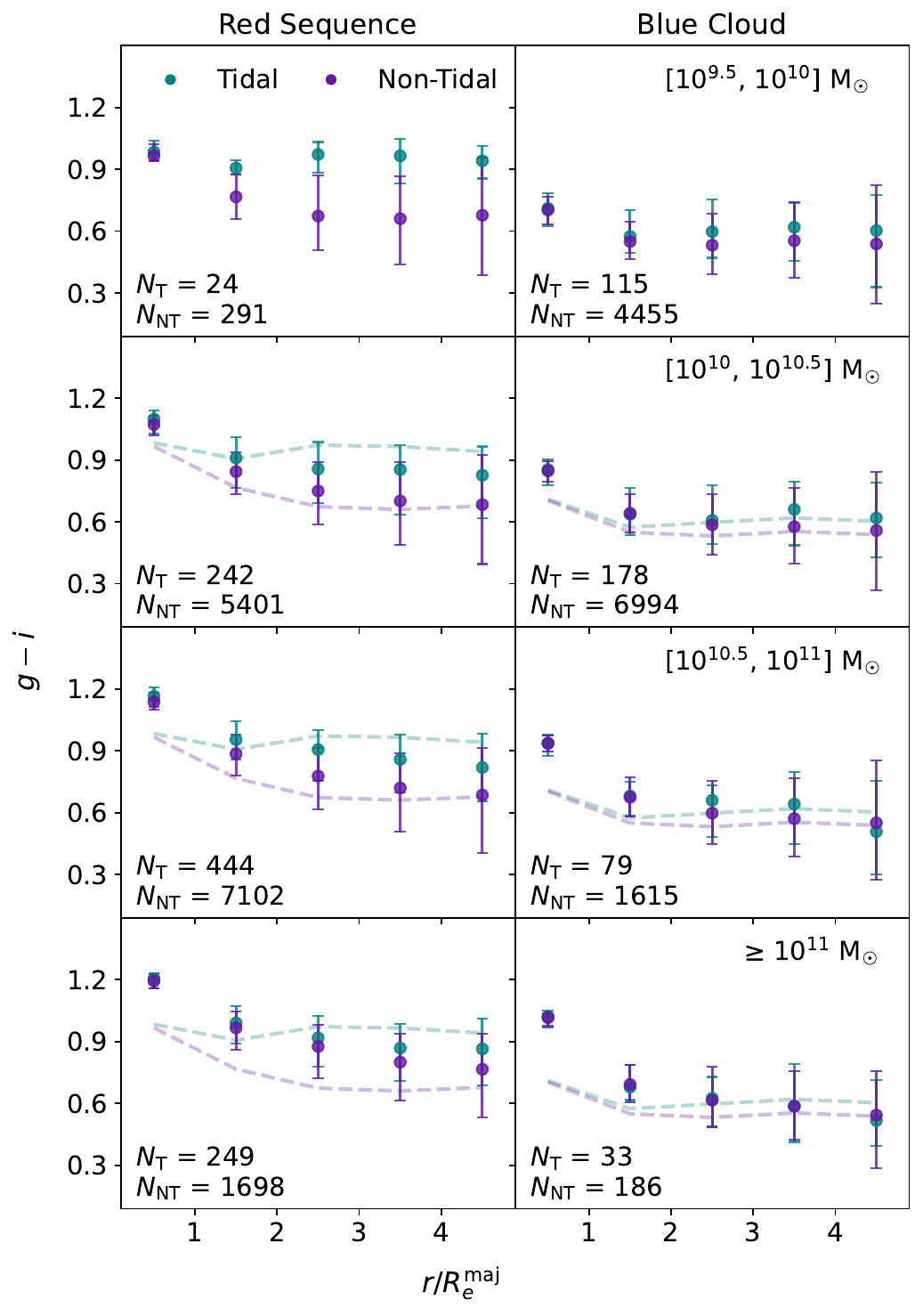}
    \caption{Distributions of radial $g-i$ colour profiles for red sequence (left) and blue cloud (right) galaxies in four stellar mass bins, given in terms of $g-i$ and the semi-major axis ($r/R_{e}^{\mathrm{maj}}$). The profiles are shown for galaxies with tidal features (teal) and without tidal feature (purple), and are measured using elliptical apertures with integer spacing from 0 to 5~$R_{e}^{\mathrm{maj}}$. Each point shows the median $g-i$ in that aperture, with the error bars depicting the $25^{\mathrm{th}}$ and $75^{\mathrm{th}}$ percentiles. The number of galaxies with and without tidal features used to generate the profiles are indicated in each panel by $N_{\mathrm{T}}$ and $N_{\mathrm{NT}}$, respectively. The dashed lines in each panel show the profiles for the lowest stellar mass bin to aid comparison between mass bins. Red sequence galaxies with tidal features tend to have redder outskirts than galaxies without tidal features, while for the blue cloud, both populations exhibit similar profiles.}
    \label{fig:colour_r_re}
\end{figure}

In Fig.~\ref{fig:colour_r_re} we compare the $g-i$ colour profiles of galaxies with and without tidal features in stellar mass bins, for both red sequence and blue cloud galaxies. At all stellar masses, both red sequence and blue cloud galaxies with and without tidal features generally exhibit radial profiles with negative gradients which tend to flatten beyond $2R_{e}^{\mathrm{maj}}$, especially for blue cloud galaxies. In the red sequence sample, at all stellar masses, we find that while the colours within $1R_{e}^{\mathrm{maj}}$ of galaxies with and without tidal features are comparable, galaxies with tidal features appear to have redder outskirts than galaxies without tidal features. Since there is significant overlap between the colour profiles, we check whether the redder colours of galaxies with tidal features are statistically significant by conducting a KS test, the results of which are presented in Table~\ref{tab:col_ks_test}. These results confirm that while red sequence tidal feature hosts tend to be redder at all radii, the difference in colour distributions become more significant beyond $2R_{e}^{\mathrm{maj}}$, the radius at which Fig.~\ref{fig:mu_i_r_re} indicated tidal features visibly contribute to a galaxy's brightness profile. We calculate the mean colour difference between galaxies with and without tidal features, $\Delta(g-i)$, for all stellar mass bins within $1R_{e}^{\mathrm{maj}}$ and beyond $2R_{e}^{\mathrm{maj}}$ using bootstrap resampling to obtain the mean colour of both distributions and the standard error on this mean. We find that within $1R_{e}^{\mathrm{maj}}$, $\Delta(g-i)=0.031\pm0.003$, while beyond $2R_{e}^{\mathrm{maj}}$, $\Delta(g-i)=0.118\pm0.007$, indicating that while the colour profiles of tidal feature hosts on the red sequence are somewhat redder than galaxies without tidal features within $1R_{e}^{\mathrm{maj}}$, this deviation significantly increases when considering the region beyond $2R_{e}^{\mathrm{maj}}$.

For blue cloud galaxies, the differences in colour distributions between galaxies with and without tidal features are much less pronounced. Fig.~\ref{fig:colour_r_re} and the KS-test results in Table~\ref{tab:col_ks_test}, indicate that the colours of tidal feature hosts and non-tidal feature hosts are consistent within $0-2R_{e}^{\mathrm{maj}}$. Beyond this radius, tidal feature hosts are only significantly redder in some radial apertures and stellar mass bins. Calculating the mean colour difference between blue cloud galaxies with and without tidal features using the same bootstrap resampling method as above we find $\Delta(g-i)=0.021\pm0.007$ within $1R_{e}^{\mathrm{maj}}$, and $\Delta(g-i)=0.036\pm0.011$ beyond $2R_{e}^{\mathrm{maj}}$. While tidal feature-hosting blue cloud galaxies do tend to have slightly redder colours than non-tidal feature host counterparts, the magnitude of this colour difference does not vary significantly with radius.

\begin{figure}
    \centering
    \includegraphics[width=0.98\columnwidth]{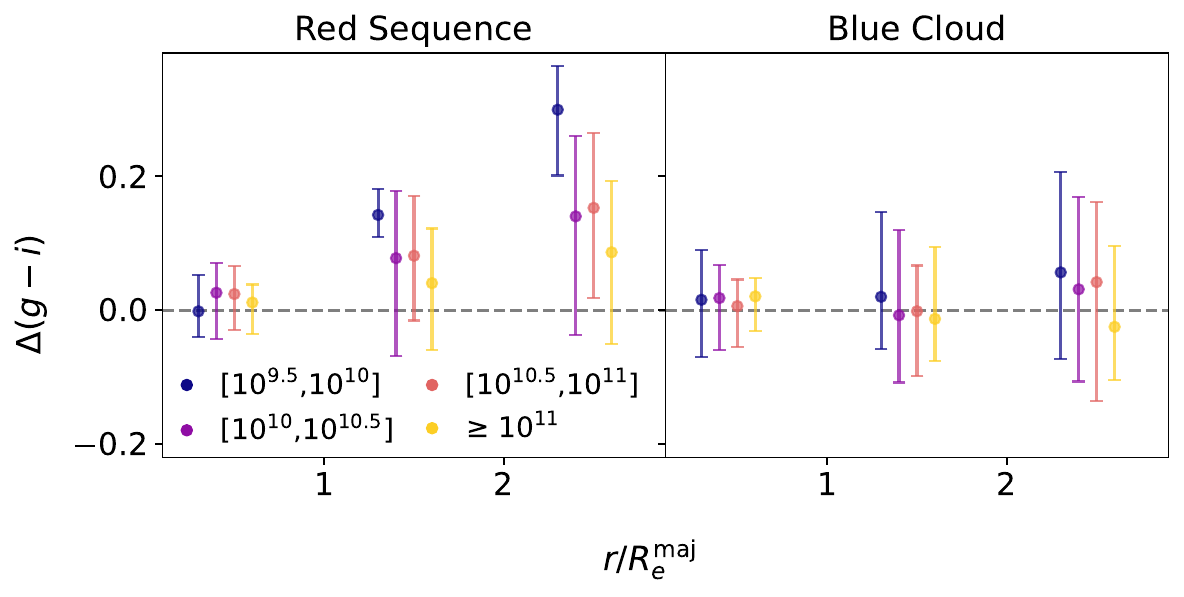}
    \caption{Radial profiles of $\Delta(g-i)$, the mean colour difference between galaxies with and without tidal features, for each of the stellar mass bins presented in Fig~\ref{fig:colour_r_re}. Each point shows the median $\Delta(g-i)$ in that aperture, with the error bars depicting the $25^{\mathrm{th}}$ and $75^{\mathrm{th}}$ percentiles. For each stellar mass bin, we sum the flux in the last 3 apertures ($2-5R_{e}^{\mathrm{maj}}$) to obtain $\Delta(g-i)$ beyond $2R_{e}^{\mathrm{maj}}$, displayed by the third series of points in each panel. The points in individual stellar mass bins are staggered for improved visibility. The lowest mass red sequence galaxies have the reddest outskirts compared to their non-tidal feature hosting counterparts.}
    \label{fig:delta_g_i_all}
\end{figure}

Fig.~\ref{fig:delta_g_i_all} depicts a simplified version of Fig.~\ref{fig:colour_r_re}, showing the median $\Delta(g-i)$ for the same stellar mass bins, and the $25^{\mathrm{th}}$ and $75^{\mathrm{th}}$ percentiles of the $\Delta(g-i)$ distributions. We combine the last 3 apertures to get the value of $\Delta(g-i)$ beyond $2R_{e}^{\mathrm{maj}}$. As with the results in Fig.~\ref{fig:colour_r_re}, we see that for red sequence galaxies, $\Delta(g-i)$ increases with radius, and is largest for galaxies with $9.5\leq\log_{10}(M_{\star}/\mathrm{M_{\odot}})<10.0$. We find that for the most massive red sequence galaxies ($\log_{10}(M_{\star}/\mathrm{M_{\odot}})\geq11.0$), $\Delta(g-i)$ is smaller than in other stellar mass bins, although this is not statistically significant. While the outskirt colours of these massive galaxies with tidal features are still redder than their non-tidal feature counterparts, their colour difference is less pronounced. The colours of blue cloud galaxies with tidal features at all radii tend to be consistent with the colours of galaxies without tidal features.

\begin{table*}
\centering
\begin{tabular}{cccccccc}
\hline
$\log_{10}(M_\star/\mathrm{M}_\odot)$ & $N_{\mathrm{T}}$ & $N_{\mathrm{NT}}$ & $0{-}1R_e$ & $1{-}2R_e$ & $2{-}3R_e$ & $3{-}4R_e$ & $4{-}5R_e$ \\ \hline
\multicolumn{8}{c}{Red Sequence} \\ \hline
$9.5{-}10.0$ & 24 & 291 & $-0.2$, $\num{6.4}{-1}$ & $-0.6$, $\ubnum{4.2}{-7}$ & $-0.6$, $\ubnum{1.5}{-7}$ & $-0.6$, $\ubnum{3.8}{-7}$ & $-0.5$, $\ubnum{1.8}{-6}$ \\
$10.0{-}10.5$ & 242 & 5401 & $-0.2$, $\ubnum{1.9}{-6}$ & $-0.2$, $\ubnum{6.0}{-8}$ & $-0.2$, $\ubnum{1.5}{-8}$ & $-0.2$, $\ubnum{2.3}{-10}$ & $-0.2$, $\ubnum{6.4}{-10}$ \\
$10.5{-}11.0$ & 444 & 7102 & $-0.2$, $\ubnum{2.9}{-12}$ & $-0.2$, $\ubnum{1.5}{-15}$ & $-0.3$, $\ubnum{9.4}{-25}$ & $-0.2$, $\ubnum{1.2}{-20}$ & $-0.2$, $\ubnum{1.1}{-18}$ \\
$11.0{-}11.7$ & 249 & 1698 & $-0.1$, $\bnum{2.0}{-2}$ & $-0.1$, $\bnum{6.9}{-3}$ & $-0.1$, $\ubnum{1.6}{-3}$ & $-0.1$, $\ubnum{3.8}{-4}$ & $-0.2$, $\ubnum{4.1}{-7}$ \\ \hline
\multicolumn{8}{c}{Blue Cloud} \\ \hline
$9.5{-}10.0$ & 115 & 4455 & $-0.1$, $\num{7.5}{-1}$ & $-0.1$, $\num{6.3}{-2}$ & $-0.2$, $\ubnum{2.2}{-3}$ & $-0.1$, $\num{7.1}{-2}$ & $0.1$, $\bnum{3.0}{-2}$ \\
$10.0{-}10.5$ & 178 & 6994 & $0.1$, $\num{1.6}{-1}$ & $-0.1$, $\num{1.1}{-1}$ & $-0.1$, $\num{1.5}{-1}$ & $-0.1$, $\bnum{4.3}{-3}$ & $-0.1$, $\ubnum{1.6}{-3}$ \\
$10.5{-}11.0$ & 79 & 1615 & $0.1$, $\num{2.4}{-1}$ & $0.1$, $\num{7.6}{-1}$ & $-0.1$, $\num{2.4}{-1}$ & $-0.2$, $\bnum{3.4}{-2}$ & $0.1$, $\num{2.7}{-1}$ \\
$11.0{-}11.7$ & 33 & 186 & $-0.1$, $\num{8.4}{-1}$ & $0.1$, $\num{9.3}{-1}$ & $0.1$, $\num{9.7}{-1}$ & $-0.1$, $\num{9.2}{-1}$ & $0.1$, $\num{9.0}{-1}$ \\ \hline
\end{tabular}
\caption{The test statistics and p-values from the two-sided KS test comparing the colour distributions of galaxies with and without tidal features, for each of the stellar mass and radial bins of red sequence and blue cloud galaxies shown in Fig.~\ref{fig:colour_r_re}. The null hypothesis is that the cumulative density function of tidal feature host and non-tidal feature hosts colours are the same. Rejection of the null hypothesis with a positive test statistic indicates that tidal feature hosts have bluer colours, while a negative test statistic indicates that tidal feature hosts have redder colours. Numbers in bold indicate $p < 0.05$ (rejection of the null hypothesis with $2\sigma$ confidence), bold+underlined numbers indicate $p < 0.003$ (rejection of the null hypothesis with $3\sigma$ confidence). Red sequence tidal feature hosts are consistently redder than non-tidal feature hosts, with the colour difference becoming more significant at radii beyond $2R_{e}^{\mathrm{maj}}$. Blue cloud galaxies show fewer distribution differences, with redder tidal feature hosts being present primarily in the lower stellar mass bins.}
\label{tab:col_ks_test}
\end{table*}

\begin{figure}
    \centering
    \includegraphics[width=0.99\columnwidth]{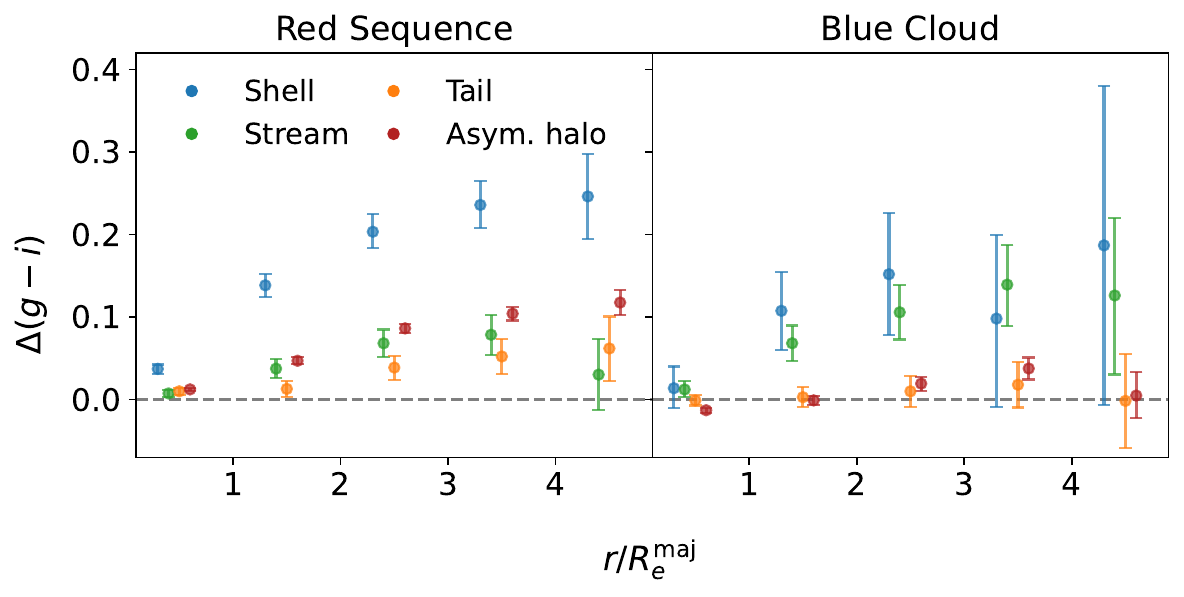}
    \caption{Radial profiles of $\Delta(g-i)$, the mean colour difference between galaxies with and without tidal features, for each type of tidal feature in our sample. Each point shows the median $\Delta(g-i)$ in that aperture, with the error bars depicting the $25^{\mathrm{th}}$ and $75^{\mathrm{th}}$ percentiles. For each tidal feature category, the contribution of galaxies without tidal features to $\Delta(g-i)$ is from a sample whose stellar mass distribution is matched to the distribution of the given feature type. The points are staggered about the centre of each aperture for improved visibility. The number of galaxies with shells, streams, tails, and asymmetric haloes are [56, 91, 111, 726] for the red sequence and [4, 22, 69, 318] for the blue cloud. Red sequence galaxies hosting shells have significantly redder outskirts than their non-tidal feature hosting counterparts.}
    \label{fig:colour_r_re_tf_types}
\end{figure}

In Fig.~\ref{fig:colour_r_re_tf_types} we separate our tidal feature sample into individual feature categories (shell, stream, tail, asymmetric halo) to determine whether the colour profiles of galaxies vary based on the types of tidal features they host. We do this by calculating $\Delta(g-i)$, the mean colour difference between galaxies with and without tidal features, in each radial bin for each tidal feature category. We must account for the variation in stellar mass distributions of galaxies hosting different types of tidal features, however, due to the small number of shells and streams in our sample, separating our sample into stellar mass bins is not possible. Instead, we use mass-matching to ensure the contribution from non-tidal galaxies to $\Delta(g-i)$ comes from a sample with the same stellar-mass distribution as the galaxies hosting a given tidal feature type. For each feature category we calculate the number of host galaxies in 9 equally-spaced stellar mass bins spanning $\mathrm{log}_{10}(M_{\star}/\mathrm{M_{\odot}})=[9.5,~11.75]$ and select identical numbers of galaxies without tidal features in each bin such that the stellar mass distributions are identical. We repeat this process 10,000 times for each feature type for both the red sequence and blue cloud galaxies, and calculate $\Delta(g-i)$ in each radial aperture for each iteration. 

Fig.~\ref{fig:colour_r_re_tf_types} shows the median and $25^{\rm{th}}$ and $75^{\rm{th}}$ percentiles of this $\Delta(g-i)$ distribution for each feature type for red sequence and blue cloud galaxies. While red sequence galaxies with any type of tidal feature tend to have redder outskirts than galaxies without tidal features, galaxies with shells have significantly redder outskirts than galaxies with any other feature type. The colour difference between red sequence galaxies without tidal features and galaxies hosting tails, streams, and asymmetric haloes are consistent. For blue cloud galaxies, the colours of tail and asymmetric halo hosts are consistent at all radii with galaxies not hosting tidal features, while hosts of shells and streams tend to be redder at radii beyond $1R_{e}^{\mathrm{maj}}$. 

\subsection{Tidal feature stellar mass}
\label{sec:feat_masses}
In this Section we estimate the stellar mass contained in tidal features. If we assume that tidal features become prominent sources of light beyond $2R_{e}^{\mathrm{maj}}$ (see Section~\ref{sec:brightness_prof}), then by subtracting the stellar mass contained within $2-5R_{e}^{\mathrm{maj}}$ in galaxies without tidal features from the mass in the same aperture in galaxies with tidal features, we should be able to isolate the stellar mass of the tidal features themselves. We sum the flux within $2-5R_{e}^{\mathrm{maj}}$ for galaxies in our sample and obtain stellar masses using the empirical relation between stellar mass, $g-i$ colour, and i-band luminosity presented in \citet{Taylor2011GAMAMassEst}. The relation takes the form $\log_{10}(M_{\star}/\mathrm{M}_{\odot})=1.15+0.70(g-i)-0.4M_i$, where $M_i$ is the absolute magnitude of the galaxy in the rest frame $i$-band. We then use bootstrapping with 10,000 resamples to obtain the mean stellar mass difference, and error on this mean, within $2-5R_{e}^{\mathrm{maj}}$ between galaxies with and galaxies without tidal features. 

Table~\ref{tab:TF_masses} shows the obtained tidal feature masses for red sequence and blue cloud galaxies in four stellar mass bins. We also calculate the ratio between tidal feature and host stellar mass for each stellar mass bin, following the same bootstrapping procedure, displayed in Table~\ref{tab:TF_masses}. We note that this is not the mass ratio of the galaxies involved in the merger but rather a lower limit on this mass ratio, as we are calculating the mass of the tidal features and not their progenitors. We find that the tidal features around red sequence galaxies are typically more massive than those around blue cloud galaxies, apart from the highest stellar mass bins in the blue cloud sample, which contains only 33 galaxies with tidal features. However, the mass ratios of tidal features to their hosts in each stellar mass bin are consistent between red sequence and blue cloud galaxies. This suggests that the higher mass of tidal features around red sequence hosts is a consequence of the mean host masses being different within identical red sequence and blue cloud stellar mass bins. While there is no visible trend between host galaxy mass and tidal feature mass in blue cloud galaxies, tidal features around red sequence galaxies tend be more massive with increasing host mass, excluding the lowest stellar mass bin which contains only 24 galaxies with tidal features. In both red sequence and blue cloud galaxies we see a decrease in mass ratio with increasing host stellar mass for $9.5\leq\mathrm{log}_{10}(M_{\star}/\mathrm{M_{\odot}})<11.0$. The fact that this is seen in both red sequence and blue cloud galaxies suggests that the relationship we observe between tidal feature mass and host mass in the red sequence may also be present in the blue cloud but hidden in the larger range of tidal feature masses for these galaxies.

\begin{table}
\renewcommand\arraystretch{1.7}
\centering
\begin{tabular}{cccc}
\hline
\multicolumn{2}{c}{$\log_{10}(M_{\star,\mathrm{H}}/\mathrm{M}_{\odot})$} & $\log_{10}(M_{\star,\mathrm{TF}}/\mathrm{M}_{\odot})$ & $M_{\star,\mathrm{TF}}/M_{\star,\mathrm{H}}$\\ \hline
\multicolumn{4}{c}{Red Sequence} \\ \hline
9.5 - 10.0 & $9.90_{0.11}^{0.09}$ & $9.23 \pm 0.08$ & $0.24 \pm 0.04$\\
10.0 - 10.5 & $10.30_{0.14}^{0.14}$ & $9.07 \pm 0.06$ & $0.12 \pm 0.01$\\
10.5 - 11.0 & $10.72_{0.16}^{0.17}$ & $9.30 \pm 0.04$ & $0.078 \pm 0.004$\\
11.0 - 11.7 & $11.16_{0.12}^{0.12}$ & $9.49 \pm 0.09$ & $0.07 \pm 0.01$\\
\hline
\multicolumn{4}{c}{Blue Cloud} \\ \hline
9.5 - 10.0 & $9.78_{0.16}^{0.16}$ & $8.69 \pm 0.09$ & $0.19 \pm 0.02$\\
10.0 - 10.5 & $10.24_{0.16}^{0.16}$ & $8.96 \pm 0.11$ & $0.13 \pm 0.02$\\
10.5 - 11.0 & $10.67_{0.14}^{0.16}$ & $8.78 \pm 0.43$ & $0.08 \pm 0.02$\\
11.0 - 11.7 & $11.15_{0.12}^{0.14}$ & $9.72 \pm 0.16$ & $0.08 \pm 0.02$\\
\hline
\end{tabular}
\caption{Stellar mass contained in tidal features, and resulting mass ratio for red sequence and blue cloud galaxies in four host stellar mass bins. The first column shows the stellar mass bins of the host galaxies, while the second column shows the mean of the stellar masses in each bin. The tidal feature masses are obtained by subtracting the stellar mass contained in the $2-5R_{e}^{\mathrm{maj}}$ aperture of galaxies without tidal features from that contained in galaxies with tidal features. The uncertainties on the host stellar masses show the $16^{\mathrm{th}}$ and $84^{\mathrm{th}}$ percentiles of the distributions, while the uncertainties in the last two columns show standard error on the mean obtained from bootstrapping. The number of galaxies in each bin are identical to those displayed in Table~\ref{tab:col_ks_test}.}
\label{tab:TF_masses}
\end{table}

Next, we calculate the mean stellar mass and mass ratios for the different morphologies of tidal features using the same method and the mass-matching and resampling method described in Section~\ref{sec:colour_prof}. For host galaxies exhibiting a given feature type, we obtain a stellar mass-matched distribution of galaxies without tidal features and calculate the difference in the mean stellar mass contained within $2-5R_{e}^{\mathrm{maj}}$ between the two samples. We repeat this 10,000 times for red sequence and blue cloud galaxies hosting each type of tidal feature. The results for each feature type for red sequence and blue cloud galaxies are presented in Table~\ref{tab:TF_type_masses}, along with the mean stellar mass of the host galaxies. Shells around red sequence galaxies have significantly higher stellar masses than any other type of feature, and higher mass ratios. In both red sequence and blue cloud galaxies, tails have the lowest mean stellar mass, significantly lower than the stellar mass contained in streams. The mass ratios of each tidal feature type around blue cloud hosts tend to be higher than those of red sequence hosts for the same feature types, apart from tails where the mass ratios are consistent. This is likely a consequence of the blue cloud host galaxies being less massive, as we found in Table~\ref{tab:TF_masses} that mass ratios tended to decrease with increasing host galaxy mass.

\begin{table*}
\renewcommand\arraystretch{1.7}
\centering
\resizebox{\textwidth}{!}{%
\begin{tabular}{ccccc|cccc}
\hline
 & \multicolumn{4}{c}{Red Sequence} & \multicolumn{4}{c}{Blue Cloud} \\ \hline
Feature type & $N_{\mathrm{T}}$ & $\log_{10}(M_{\star,\mathrm{H}}/\mathrm{M}_{\odot})$ & 
$\log_{10}(M_{\star,\mathrm{TF}}/\mathrm{M}_{\odot})$ &
$M_{\star,\mathrm{TF}}/M_{\star,\mathrm{H}}$ & $N_{\mathrm{T}}$ & $\log_{10}(M_{\star,\mathrm{H}}/\mathrm{M}_{\odot})$ & 
$\log_{10}(M_{\star,\mathrm{TF}}/\mathrm{M}_{\odot})$ &
$M_{\star,\mathrm{TF}}/M_{\star,\mathrm{H}}$\\ \hline
Shell & 56 & $10.78_{0.59}^{0.52}$ & $9.66 \pm 0.09$ & $0.08 \pm 0.01$ & 4 & $10.12_{0.20}^{0.20}$ & $9.26 \pm 0.21$ & $0.15 \pm 0.04$\\
Stream & 91 & $10.68_{0.37}^{0.45}$ & $9.31 \pm 0.14$ & $0.05 \pm 0.01$ & 22 & $10.48_{0.40}^{0.47}$ & $9.51 \pm 0.10$ & $0.11 \pm 0.02$\\
Tail & 111 & $10.72_{0.34}^{0.38}$ & $8.94 \pm 0.28$ & $0.02 \pm 0.01$ & 69 & $10.28_{0.45}^{0.43}$ & $8.74 \pm 0.27$ & $0.03 \pm 0.01$\\
Asym. halo & 726 & $10.76_{0.38}^{0.36}$ & $9.15 \pm 0.07$ & $0.025 \pm 0.004$ & 318 & $10.29_{0.42}^{0.47}$ & $9.07 \pm 0.05$ & $0.06 \pm 0.01$ \\ \hline
\end{tabular}
}
\caption{Number of tidal features, mean host stellar mass, stellar mass contained in the different tidal feature types, and the resulting mass ratios for red sequence and blue cloud galaxies. The tidal feature masses are obtained by subtracting the stellar mass contained in $2-5R_{e}^{\mathrm{maj}}$ of mass-matched galaxies without tidal features from that contained in galaxies with tidal features. The uncertainties on the host stellar masses show the $16^{\mathrm{th}}$ and $84^{\mathrm{th}}$ percentiles of the distributions, while the uncertainties in the tidal feature mass and mass ratio columns show standard error on the mean obtained from bootstrapping.}
\label{tab:TF_type_masses}
\end{table*}

\section{Discussion}
\label{sec:disc}
In this work, we have used data from the HSC-SSP survey to model and measure the radial surface brightness and $g-i$ colour profiles of 31,819 galaxies, including 1415 galaxies with tidal features. We compared the colour profiles of galaxies with and without tidal features for both the full sample of tidal features and individual tidal feature classes, and estimated the stellar mass contained in tidal features. In this section, we compare our results with those obtained from previous analyses, and investigate whether the differences in the colour profiles of galaxies with and without tidal features reveal information about the mergers responsible for the tidal features. 

\subsection{Tidal feature brightness, colour, and stellar mass}
\label{sec:disc_general}

In Section~\ref{sec:brightness_prof} we compared the surface brightness of galaxies with and without tidal features, finding brighter outskirts in galaxies with tidal features beyond $2R_{e}^{\mathrm{maj}}$. This trend was present for blue cloud galaxies at all stellar masses, but for red sequence galaxies, the deviation in surface brightness of galaxies with tidal features decreased with increasing stellar mass, and was no longer present for the most massive ($\log_{10}(M_{\star}/\mathrm{M}_{\odot})>11.0$) red sequence galaxies. We postulate that this could be the case if these massive galaxies were surrounded by more diffuse light, such as intracluster light, from being located in cluster environments, especially if they were the central galaxies of these clusters (e.g. \citealt{Mihos2005ICL}). However, we find no difference in the rates of galaxies in clusters between galaxies with and without tidal features across the different stellar mass bins. Another possible explanation is that the lack of deviation in surface brightness is a direct consequence of the two-phase model of galaxy formation \citep{Oser2010GalFormationPhases}, wherein since $z<2$ the mass growth of galaxies is dominated by accretion through mergers depositing stellar material in the outkirts of galaxies. The importance of accretion increases with host galaxy stellar mass, with accreted stars accounting for up to 87 per cent of the stellar mass of the most massive galaxies \citep{Oser2010GalFormationPhases}. Hence, the more massive red sequence galaxies, despite not exhibiting obvious tidal features, likely have richer merger histories than lower mass galaxies, resulting in their outskirts having elevated levels of light from these past mergers.

This brings up an important factor to consider when interpreting the results of our analysis. As a result of our methodology relying on measuring the light in apertures, our colour and brightness measurements do not only contain contributions from the tidal features, but also from phase-mixed accreted material from past merger events. The sample of galaxies without identifiable tidal features does not indicate a sample free of mergers, but instead the tidal feature sample traces a subset of galaxies with more recent mergers, or mergers with more significant accretion, resulting in observable tidal features. While our comparison of the sample of galaxies with tidal features to one without tidal features should in part isolate the contribution from tidal features, the observed surface brightness and colour differences will also contain contributions from any differences in accretion histories between the two samples.

In Section~\ref{sec:colour_prof} we analysed the $g-i$ colour profiles of galaxies in our samples, finding negative colour gradients for both red sequence and blue cloud galaxies. This is in agreement with the observational work of \citet{Gonzalez-Perez2011SDSS_col_grads}, based on SDSS DR7 \citep{Abazajian2009SDSS_DR7} data, who found that both early- and late-type galaxies had cores which were redder than their outskirts, when looking out to 2 half-light radii. This is also in agreement with the work of \citet{Khalid2025TFColours}, who analysed the colour profiles of simulated galaxies in mock images, and found negative gradients for red sequence galaxies. However, they found significant scatter and no visible trend in the colour profiles of blue cloud galaxies. The negative $g-i$ colour gradients observed in red sequence galaxies is consistent with the light in their outskirts being composed of light accreted through predominantly minor and mini mergers (e.g. \citealt{Oser2012ETG_accretion, Remus2022MagneticumAccMass}). 

When comparing the colours of galaxies with and without tidal features (Figs.~\ref{fig:colour_r_re} and \ref{fig:delta_g_i_all}), we found that red sequence galaxies with tidal features had significantly redder outskirts than non-tidal feature hosts. Beyond $2R_{e}^{\mathrm{maj}}$, we found a mean colour difference between tidal feature and non-tidal feature hosts $\Delta(g-i)=0.118\pm0.007$. In blue cloud galaxies, we found that tidal feature hosts were redder than non-tidal feature hosts only in certain mass bins and radial apertures, but overall, tidal feature hosts showed no significant colour variation from non-tidal feature hosts. These results contradict those of \citet{Khalid2025TFColours}, who find a slight tendency for red sequence tidal feature hosts with $10.0\leq\mathrm{log}_{10}(M_{\star}/\mathrm{M_{\odot}})<11.0$ to have bluer outskirts. However, their results show scatter between the three different hydrodynamical simulations compared, suggesting that these trends may be dominated by the different subgrid physics of the simulations. \citet{Yoon2023ETG_merger_profiles} analysed the age and metallicity gradient of 193 early-type galaxies with $M_\mathrm{\star}\geq10^{10}\mathrm{M_{\odot}}$, 44 of which had tidal features, using Mapping Nearby Galaxies at Apache Point Observatory (MANGA; \citealt{Bundy2015MANGA}) IFU observations and SDSS Stripe 82 \citep{Fliri2016SDSS82} images. They found that the 44 galaxies exhibiting tidal features had lower metallicities and younger ages than non-tidal feature hosts. This disagrees with our findings as lower metallicities and younger ages should correspond to bluer colours. One possible explanation is that \citet{Yoon2023ETG_merger_profiles} limited their analysis out to $1.5R_{e}$, whereas we observe the most significant colour differences between tidal feature and non-tidal feature hosts beyond $2R_{e}^{\mathrm{maj}}$. \citet{KadoFong2018HSCTidalFeat} used a sample of galaxies drawn from the Wide layer of the HSC-SSP to analyse the $g-i$ colours of tidal features, and found tidal features to be generally bluer than the core of their host galaxies. While their methodology differs from ours in that they isolate the individual tidal features as opposed to considering radial apertures, this result agrees with our findings as the negative colour gradients we observed for galaxies with tidal features indicate that the galaxy outskirts where the tidal features are located are indeed bluer than the core of the host galaxies. However, these negative colour gradients were also observed in galaxies without tidal features, indicating that they could be attributed to light accreted through older merger events, rather than the mergers responsible for the observed tidal features.

\citet{Remus2022MagneticumAccMass} used the Magneticum hydrodynamical cosmological simulation to study the merger histories of different classes of galaxies, based on their distribution of accreted and in-situ components. They found that the most common class (72 per cent) of galaxies were accretion dominated at large radii, and that a significant portion of the accreted mass originated from gas-poor minor and mini mergers. This could explain the trends we observe for the red sequence, as the bluer colour of the outskirts of tidal feature hosts compared to their cores would be expected from minor mergers with less massive satellites, while the redder outskirt colours of galaxies with tidal features compared to non-tidal feature hosts may reflect the gas-poor nature of these mergers.

Many works, both based on observational and simulation data, find evidence that mergers can induce star formation (SF) in galaxies, with merging and post-merger galaxies exhibiting enhanced star formation rates (SFRs; e.g. \citealt{Pearson2019MergersSFR, Bottrell2024MiniMergersSFAsym, Ferreira2025MergerSFMassGrowth}) or higher rates of starbursts (e.g. \citealt{Ellison2013PairsStarburst, Silva2018MergerStarburst, Gordon2025TF_SF}) and post-starbursts (e.g. \citealt{Pawlik2018PostStarburstMergers, Ellison2022PostMergerPostStarburst, Wilkinson2022MergersStarburst, Gordon2025TF_SF}) compared to non-merging galaxies. Since higher SFRs are associated with bluer colours, we would expect these to appear as bluer colours in either the inner or outer regions of galaxies with tidal features compared to galaxies without tidal features. However, we do not find this to be the case in our red sequence or blue cloud galaxies. One possible explanation for this is that while SFR enhancements and starbursts can often have merger origins, the fraction of mergers that cause SFR enhancements and starbursts compared to the overall merger population is still relatively low, ranging from $\sim5-20$ per cent (e.g. \citealt{Ellison2022PostMergerPostStarburst, Gordon2025TF_SF}). This means that while there are likely SFR-enhanced and starburst galaxies in our tidal feature sample, they may just be hidden in the scatter of the tidal feature colour distribution. In particular, this may explain why we do not see significantly redder outskirts in blue cloud tidal feature hosts, as any merger-induced SFR enhancements are likely to occur in these gas-rich galaxies rather than in the gas-poor red sequence sample. Another factor to consider regarding works based on simulation data that find bluer colours or enhanced SFRs in galaxies with tidal features (e.g. \citealt{Bottrell2024MiniMergersSFAsym, Khalid2025TFColours}) is the effect of the simulation subgrid physics. \citet{Sparre2016IllustrisZoom} showed that the resolution of the Illustris simulation was insufficient to resolve starburst which could trigger quenching through stellar feedback. This could cause higher rates of merger-induced SFR enhancements, and therefore bluer colours, than are present in observational data. \citet{Omori2025Mergers_SF_AGN} analysed the star formation rates of $\sim100$ star-forming post-merger galaxies using HSC-SSP imaging and GAMA spectra to obtain galaxy properties. In contrast to other works, they found that post-mergers displayed signs of suppressed star formation. We find no evidence of this in our sample of blue cloud galaxies, but as per our previous argument, it could be present within the scatter. Alternatively, our lack of observed star formation enhancement or suppression could be attributed to limitations in our method, as colour on its own may not be sensitive enough as a tracer of star formation.

In Section \ref{sec:feat_masses} we calculated the stellar mass contained in tidal features and resulting mass ratios for red sequence and blue cloud galaxies in different host stellar mass bins. We found that the mean mass ratios of tidal feature to host galaxy were consistent between red sequence and blue cloud galaxies, and decreased with increasing host stellar mass from  $9.5\leq\mathrm{log}_{10}(M_{\star}/\mathrm{M_{\odot}})<11.0$. This suggests that less massive host galaxies are more likely to merge with companions which are relatively closer in mass, compared to more massive galaxies. This is likely a consequence of less massive galaxies being much more common than high mass galaxies in the Universe, a natural consequence of hierarchical structure formation (e.g. \citealt{Press1974FormGalCluster, White1978GalForm}).

\subsection{Individual feature classes}
\label{sec:disc_tf_types}
In Section~\ref{sec:colour_prof}, we compared the radial colour differences of galaxies with and without tidal features for the individual feature types. In this Section, we investigate whether the results of Fig. 10 suggest a connection between the $\Delta(g-i)$ measurements of different classes of tidal features and the properties of mergers that created them. We acknowledge however, that other galaxy properties, such as the host stellar masses and the tidal feature to host mass ratios presented in Table 4, may also play a role in the $\Delta(g-i)$ measurements obtained. Due to our measurement method relying on apertures, rather than isolating only the light associated with tidal features, our colour measurements are also impacted by light from the host galaxy. While we have minimised the impact of this light on our $\Delta(g-i)$ measurements by incorporating the contribution of galaxies without tidal features, comparing stellar mass-matched samples of each tidal feature type, and correcting for host galaxy morphology in the first order by separating red sequence and blue sequence galaxies, the trends observed are likely still impacted by the properties of the host galaxies and light in the outer regions of the galaxies not associated with the tidal features.

For the red sequence, we found that galaxies hosting shells had the reddest outskirts of any feature type, and that these galaxies showed significant colour differences from non-tidal feature hosts even within $2R_{e}^{\mathrm{maj}}$. The differences in shell host colour profiles being significant at lower radii could be an indication of shell-forming mergers impacting the host galaxy more globally, instead of being limited to the outskirts such as for other feature types. This aligns with shells being formed primarily from radial mergers with low angular momentum (e.g. \citealt{Quinn1984ShellsEllipGals, Hendel2015TidalDebOrbit, Pop2018ShellsIllustris, Stoiber2025StreamShellsMass}) which can significantly impact the dynamics of the host galaxy, such as causing them to lose their global angular momentum \citep{Valenzuela2024_TF_kins_form_hist}.

Galaxies hosting streams had redder outskirts than galaxies without tidal features in both red sequence and blue cloud galaxies, although this colour difference was less significant than that of shells. This is consistent with the work of \citet{KadoFong2018HSCTidalFeat}, who analysed 78 shells and 497 streams and found that streams had bluer colours than shells with respect to their hosts. Our results are also consistent with those of \citet{Stoiber2025StreamShellsMass}, who analysed the properties of 24 shell-hosting and 66 stream-hosting galaxies using mock images created using the Magneticum Pathfinder Simulations. Their sample was limited to $z=0.07$ and $\log_{10}(M_{\star}/\mathrm{M}_{\odot})\geq10.0$, and contained mostly ETGs. They found that shells were on average more metal-rich than streams, consistent with the redder colours we observe in shells. The tendency for stream hosts to have outskirt colours that differ from non-tidal feature hosts in both red sequence and blue cloud galaxies is consistent with streams being composed of material from satellite galaxies. Furthermore, the tendency for stream hosts to have smaller values of $\Delta(g-i)$ than shells (considering mainly the red sequence shell hosts due to the very small number of blue cloud shell hosts) suggests that the material in streams originates from satellites with lower masses than those responsible for shells. This is supported by our Section \ref{sec:feat_masses} findings, where we calculated the mass contained in the different tidal feature classes and found that shells around red sequence galaxies were significantly more massive than streams and had higher mass ratios. These findings are consistent with the literature, which attributes streams to minor mergers (e.g. \citealt{Mancillas2019ETGS_merger_hist, Sola2022TailsvsStreams, Stoiber2025StreamShellsMass}), while shells can have either minor or major merger origins (e.g. \citealt{KadoFong2018HSCTidalFeat, Pop2018ShellsIllustris, Stoiber2025StreamShellsMass}). Another possible contribution to the smaller values of $\Delta(g-i)$ found in streams compared to shells could be from the morphology of their progenitors. \citet{Stoiber2025StreamShellsMass} found that $\sim28$ per cent of streams in their sample had disky progenitors, compared to only $\sim19$ per cent of shells. While this is likely related to shell progenitors being more massive and hence more likely to be spheroidal galaxies, it would also contribute to the outskirt colours of stream hosts being bluer than those of shells.

There are limited works studying the mass of tidal features, however, \citet{Sola2025Strrings_streams} used images from the DESI Legacy Imaging Surveys to model 35 streams and calculated their stellar masses using an $r-$band mass-to-light ratio. They found that streams in their sample have a median stellar mass of $\log_{10}(M_{\star}/\mathrm{M}_{\odot})\sim8.8$. While the streams in our sample are significantly more massive with mean stellar masses $\log_{10}(M_{\star}/\mathrm{M}_{\odot})=9.31_{-0.11}^{+0.12}$ for red sequence hosts, and $\log_{10}(M_{\star}/\mathrm{M}_{\odot})=9.51_{-0.07}^{+0.08}$ for blue cloud hosts, our stream masses do agree with the higher end of their stream stellar mass distribution (Fig. 7 in \citealt{Sola2025Strrings_streams}). \citet{Sola2025Strrings_streams} also calculated the mass ratio of the streams to their hosts and found a mean ratio of 0.02. While this is a lower mass ratio than we find for streams around red sequence  ($0.05\pm0.01$) and blue cloud ($0.11\pm0.02$) stream hosts, they found four streams with mass ratios higher than 0.05. The discrepancy both in mass of streams and mass ratios could be attributed to their galaxies being at lower redshifts ($z<0.02$) and their images being better optimised for visual classification, likely allowing them to detect fainter streams.

\citet{Stoiber2025StreamShellsMass} also calculated the stellar mass of tidal features in their sample, along with the mass of their progenitors. They found a median shell mass $M_{\star}=8_{-2}^{+6}\times10^{9}\mathrm{M}_{\odot}$ and stream mass $M_{\star}=1_{-5}^{+7}\times10^{9}\mathrm{M}_{\odot}$. Comparing with our red sequence sample, given that the \citet{Stoiber2025StreamShellsMass} is mainly composed of ETGs, we find mean shell mass $M_{\star}=5_{-1}^{+1}\times10^{9}\mathrm{M}_{\odot}$ and stream mass $M_{\star}=2_{-1}^{+1}\times10^{9}\mathrm{M}_{\odot}$, which is in excellent agreement. \citet{Stoiber2025StreamShellsMass} also calculated the mass ratio of shells and streams to their progenitors, finding that the mass of shells and streams correspond to $\sim19$ and $\sim26$ per cent of the stellar mass of their progenitors, respectively. If we use these values to convert our mass ratios of tidal features to host into mass ratios of progenitor to host, we obtain mass ratios $\sim0.42$ and $\sim~0.19$ for shell and stream progenitors, respectively. While these are simple approximations, they support the findings of \citet{Stoiber2025StreamShellsMass} that shell-forming mergers have higher mass ratios $0.2<M_{\mathrm{progenitor}}/M_{\mathrm{host}}<1$, while stream-forming mergers are mostly minor and mini mergers with $M_{\mathrm{progenitor}}/M_{\mathrm{host}}<0.2$.

With regards to tails, we found that the outskirt colours of their hosts were redder than those of red sequence galaxies without tidal features, however, in blue cloud galaxies, the colour of galaxies hosting tails were consistent with non-tidal feature hosts at all radii. The colour of galaxies hosting tails being aligned with their non-tidal feature hosting counterparts in blue cloud galaxies but not in red sequence galaxies is interesting. As tails are thought to comprise of material pulled out from the host galaxy, one would expect their colours to be consistent with those of non-tidal feature hosting galaxies. One possible explanation for this inconsistency could be the difference in radial extent between red sequence and blue cloud galaxies. For a given stellar mass, blue cloud galaxies tend to have larger effective radii than red sequence galaxies (e.g. \citealt{Kawinwanichakij2021HSCMassSize, Valenzuela2024_TF_kins_form_hist}), indicating a lower concentration of light, meaning that the outskirts of blue cloud galaxies may contain more light than their red sequence counterparts. If this is the case, then the contribution of non-tidal feature blue cloud galaxies to $\Delta(g-i)$ at larger radii would be from host galaxy light rather than the background, as may be the case for red sequence galaxies. Additionally, the colour profiles of blue cloud galaxies flatten at lower radii than red sequence ones (Fig.~\ref{fig:colour_r_re}). This means that if a portion of the stellar material in the tails was displaced from regions closer to the centre of the galaxy, $\Delta(g-i)$ would be more positive for red sequence galaxies. In a study of tidal features using images from the from the Canada-France-Hawaii Telescope (CFHT), \citet{Sola2022TailsvsStreams} found that tails had bluer $g-r$ colour than streams. Similarly, in work modelling 9 streams and 5 tails using Wendelstein imaging data, \citet{Pippert2025TailStreamModel} found tails to have bluer $g-r$ colours than streams. This is consistent with our results for blue cloud galaxies, for which stream hosts had colours significantly redder than their non-tidal feature hosting counterparts compared to tails. We also find this to be the case in red sequence galaxies within $4R_{e}^{\mathrm{maj}}$, albeit very weakly.

In both red sequence and blue cloud galaxies, tails had the lowest mean stellar mass, significantly lower than then stellar mass contained in streams. This is consistent with tails being composed of material from the host galaxy in the sense that tails would be the result of stars from the host galaxy being redistributed. Hence, they would contain less excess stellar mass relative to galaxies without tidal features than streams, which are composed of material originating from a satellite galaxy. The mass ratio of tails being consistent between red sequence and blue cloud hosts despite the different mean host stellar masses can also be attributed to them being composed of material from the host galaxy. In the case of tails, the mass ratio $M_{\star,\mathrm{TF}}/M_{\star,\mathrm{H}}=3\pm1$ per cent is an indication of the amount of stellar material that has been displaced from the inner regions of the host galaxy to beyond $2R_{e}^{\mathrm{maj}}$.

\section{Conclusions}
\label{sec:conc}
In this work, we have measured the radial $g-i$ colour profiles of 31,819 galaxies drawn from the HSC-SSP optical imaging survey, including 1415 exhibiting tidal features. We compared the colour profiles of galaxies with and without tidal features to draw conclusions about the mergers that created them, including an analysis of the differences between distinct classes of tidal features and an estimation of the stellar mass contained in tidal features. These are our key findings:
\begin{enumerate}
    \item Both galaxies with and without tidal features had outskirts bluer than their cores (Fig.~\ref{fig:colour_r_re}). Over the full stellar mass range, red sequence galaxies exhibiting tidal features had redder outskirts than their non-tidal feature hosting counterparts. This consistent with the outskirts of these galaxies being dominated by stars accreted from gas-poor minor mergers.
    \item The mass ratio of tidal features to their hosts was consistent between red sequence and blue cloud galaxies, and decreased with increasing stellar mass (Table~\ref{tab:TF_masses}). This suggests that less massive galaxies undergo mergers with companions closer in mass, compared to more massive galaxies.
    \item Galaxies with shells had the reddest outskirt colours of any tidal feature class (Fig.~\ref{fig:colour_r_re_tf_types}), when compared to non-tidal feature galaxies of equivalent stellar mass. Shells around red sequence galaxies were also the most massive and had the highest mass ratios of any of the tidal feature classes (Table~\ref{tab:TF_type_masses}).
    \item Galaxies with streams had redder outskirts than galaxies without tidal features, but bluer than those of shells. This, and their lower mass ratios compared to shells, is consistent with streams being formed from mergers with galaxies less massive than those responsible for shells.
    \item Blue cloud galaxies with tails had colours consistent with material being pulled out of the inner regions of galaxies. The stellar masses of tails around both red sequence and blue cloud hosts accounted for $\sim3$ per cent of the mass of their hosts being displaced from the inner regions ($<2R_{e}^{\mathrm{maj}}$) to their outskirts.
    \item We found no evidence for star formation enhancement in galaxies with tidal features but theorise this may be due to the sensitivity of our method or present within the scatter of our results.
\end{enumerate}

We found overall agreement with past observational works analysing the colour of tidal features despite using different methods. Our findings also aligned with the predictions of merger properties from simulations, including the properties of mergers responsible for distinct tidal feature classes. This confirms the validity of our methodology and highlights that the colours of tidal features are a valuable probe into the process through which galaxies evolve.

\section*{Acknowledgements}
\label{sec:acknowledge}
This research made use of the ``K-corrections calculator'' service available at \url{http://kcor.sai.msu.ru/}. We acknowledge funding support from LSST Corporation Enabling Science grant LSSTC 2021-5. SB acknowledges funding support from the Australian Research Council through a Discovery Project DP190101943. 
The Hyper Suprime-Cam (HSC) collaboration includes the astronomical communities of Japan and Taiwan, and Princeton University. The HSC instrumentation and software were developed by the National Astronomical Observatory of Japan (NAOJ), the Kavli Institute for the Physics and Mathematics of the Universe (Kavli IPMU), the University of Tokyo, the High Energy Accelerator Research Organization (KEK), the Academia Sinica Institute for Astronomy and Astrophysics in Taiwan (ASIAA), and Princeton University. Funding was contributed by the FIRST program from the Japanese Cabinet Office, the Ministry of Education, Culture, Sports, Science and Technology (MEXT), the Japan Society for the Promotion of Science (JSPS), Japan Science and Technology Agency (JST), the Toray Science Foundation, NAOJ, Kavli IPMU, KEK, ASIAA, and Princeton University. 
This paper makes use of software developed for Vera C. Rubin Observatory. We thank the Rubin Observatory for making their code available as free software at http://pipelines.lsst.io/.
This paper is based on data collected at the Subaru Telescope and retrieved from the HSC data archive system, which is operated by the Subaru Telescope and Astronomy Data Center (ADC) at NAOJ. Data analysis was in part carried out with the cooperation of Center for Computational Astrophysics (CfCA), NAOJ. We are honored and grateful for the opportunity of observing the Universe from Maunakea, which has the cultural, historical and natural significance in Hawaii.
GAMA is a joint European-Australasian project based around a spectroscopic campaign using the Anglo-Australian Telescope. The GAMA input catalogue is based on data taken from the Sloan Digital Sky Survey and the UKIRT Infrared Deep Sky Survey. Complementary imaging of the GAMA regions is being obtained by a number of independent survey programmes including GALEX MIS, VST KiDS, VISTA VIKING, WISE, Herschel-ATLAS, GMRT and ASKAP providing UV to radio coverage. GAMA is funded by the STFC (UK), the ARC (Australia), the AAO, and the participating institutions. The GAMA website is https://www.gama-survey.org/\\ 
\section*{Data Availability}
\label{sec:data_avail}
 
All data used in this work is publicly available at \url{https://hsc-release.mtk.nao.ac.jp/doc/} (HSC-SSP data) and \url{https://www.gama-survey.org/} (GAMA data).


\bibliographystyle{mnras}
\bibliography{bibs} 


\appendix

\section{Effect of scattered light subtraction}
\label{sec:app_star_sub}

\begin{figure}
    \centering
    \includegraphics[width=0.99\columnwidth]{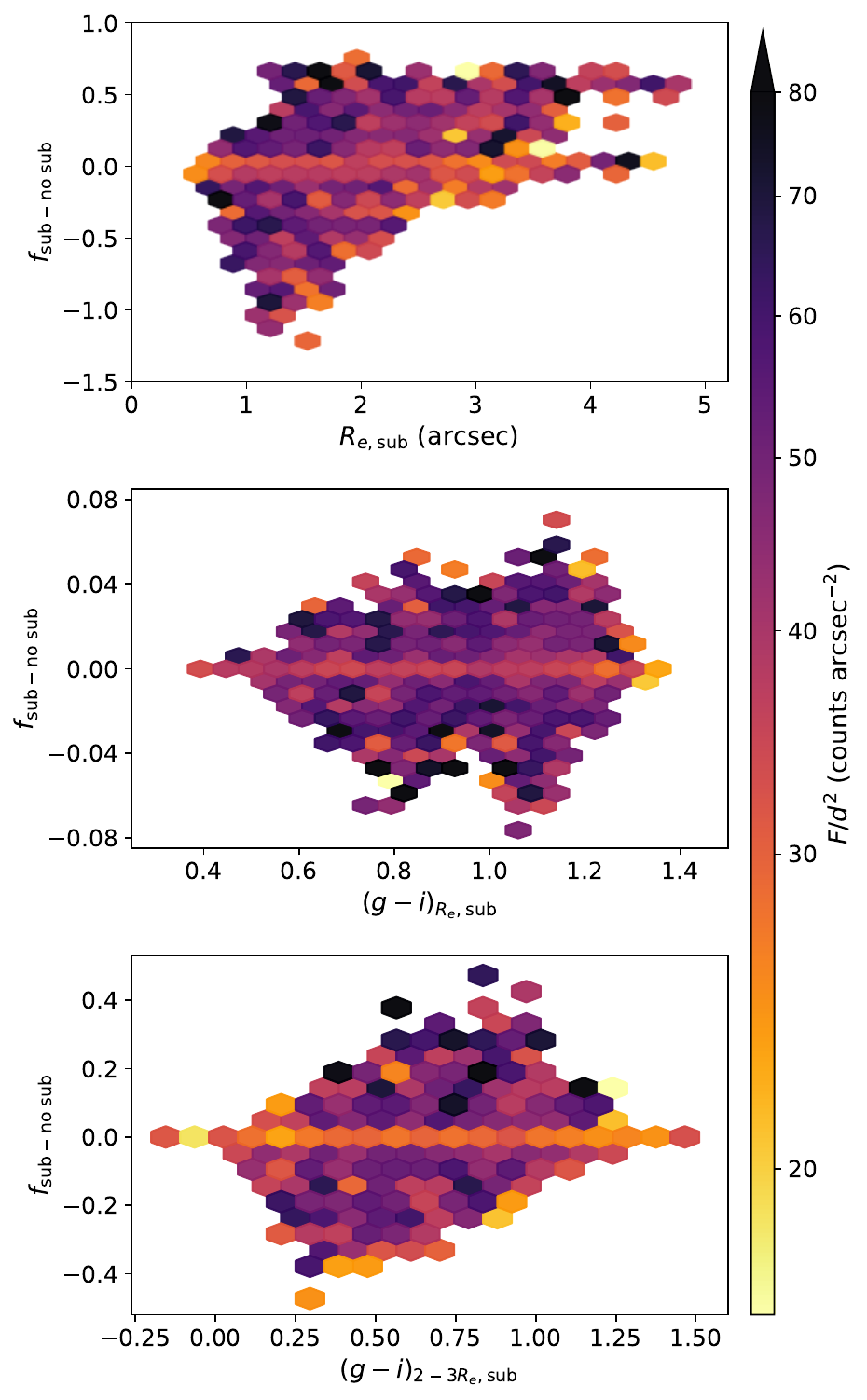}
    \caption{Distribution of measured effective semi-major radii (top), $g-i$ colour within $1R_{e}$ (middle), and $g-i$ colour within $2-3R_{e}$ with star subtraction, and their respective residuals when the measurements are made without star subtraction. The residuals (y-axis) are shown in terms of fractional deviation from the measurements obtained with star subtraction. The results are displayed in hexagonal bins and bins with fewer than 10 galaxies are not shown. The bins are coloured by the mean $F/d^{2}$, a proxy for the expected decay in the intensity of flux of a star. For a star affecting a given galaxy, $F$ is the flux of the star, and $d$ is its projected distance from the galaxy in arcseconds. For a galaxy affected by multiple stars, we sum the $F/d^{2}$ of each star.}
    \label{fig:rad_col_residuals}
\end{figure}

In this Section we analyse the impact of subtracting the scattered light from stars on our galaxy size and colour measurements. We also verify whether the star subtraction affects the results presented in Section~\ref{sec:colour_prof}, regarding the radial $g-i$ colour profiles of galaxies with and without tidal features. Fig.~\ref{fig:rad_col_residuals} compares the measured radius, $g-i$ colour within $1R_{e}^{\rm{maj}}$, and $g-i$ colour within $2-3R_{e}^{\rm{maj}}$ of the 25,770 galaxies affected by scattered light from stars, when star subtraction is and is not applied. The bins are coloured by the mean $F/d^{2}$, where $F$ is the flux of a star affecting a galaxy, and $d$ is its projected distance from a galaxy. This metric serves as a proxy for the expected decay in the intensity of flux of a star. The radii of galaxies with smaller extent on the sky tend to be more strongly affected, with some overestimated by more than 100 per cent when star subtraction is not performed. The radii of galaxies with larger extent tend to be underestimated when star subtraction is not performed. These deviations tend to increase for larger values of $F/d^{2}$, which is expected as brighter, more nearby stars will contribute more contaminating light to galaxies. The deviation of the $g-i$ colour of galaxies within $1R_{e}^{\rm{maj}}$ is symmetrical, indicating that when not performing star subtraction, galaxies are equally likely to be measured as redder or bluer, however, the scatter in the measured colour increases for redder galaxies. For colours within $1R_{e}^{\rm{maj}}$ the scatter is relatively small, with maximum deviations of $\sim8$ per cent, however this scatter increases significantly for colours measured within $2-3R_{e}^{\rm{maj}}$, with deviations as large as $\sim50$ per cent. This is expected as the fainter light in the outskirts of galaxies is more impacted by the scattered light from stars than the light in the central regions. 

\begin{figure}
    \centering
    \includegraphics[width=0.99\columnwidth]{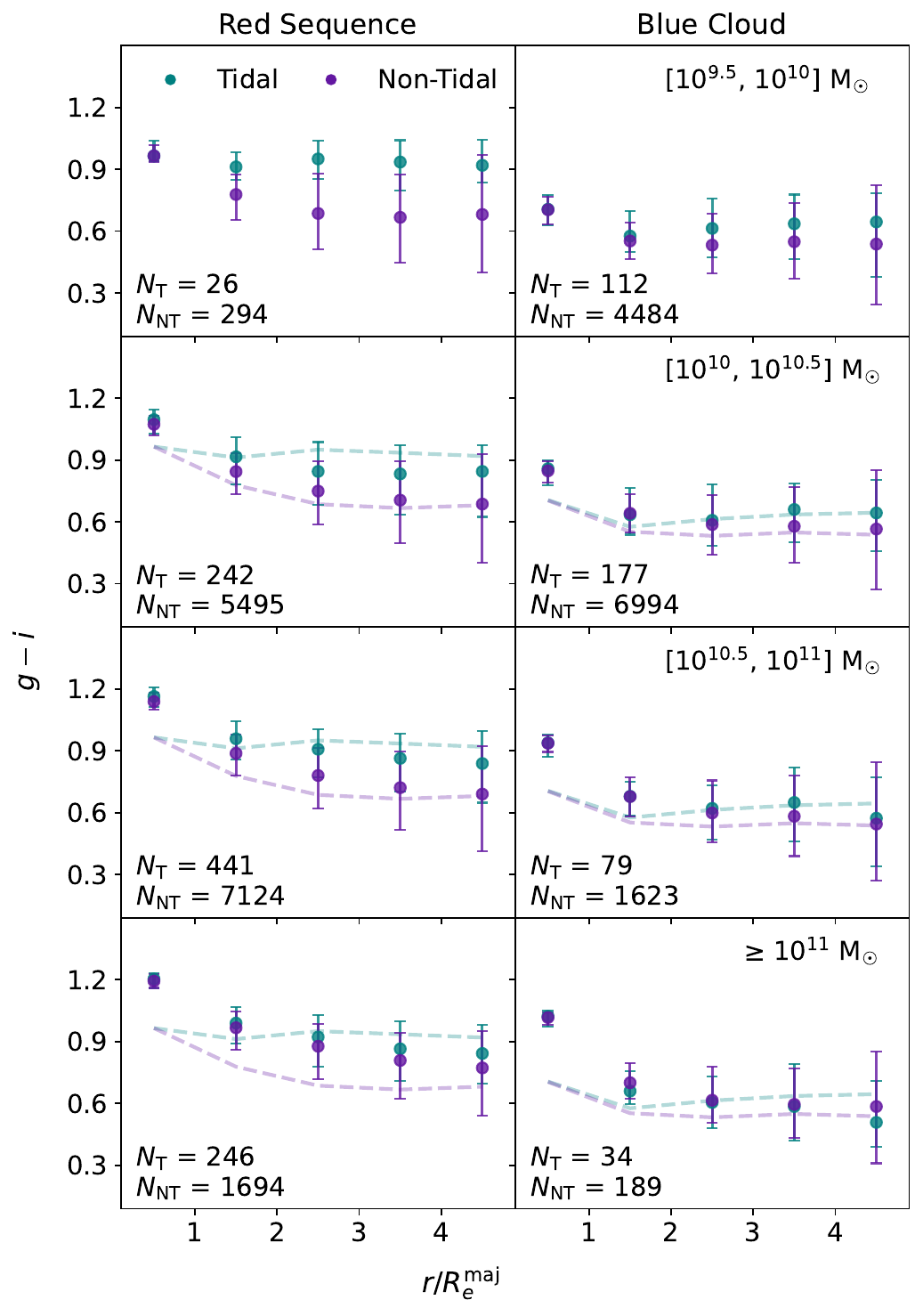}
    \caption{Distributions of radial $g-i$ colour profiles when star subtraction is not performed. The binning and display of the distributions is identical to Fig.~\ref{fig:colour_r_re}. The results remain unchanged from those presented in Fig.~\ref{fig:colour_r_re}}.
    \label{fig:colour_r_re_no_star_sub}
\end{figure}

Fig.~\ref{fig:colour_r_re_no_star_sub} is a version of Fig.~\ref{fig:colour_r_re}, showing the $g-i$ colour profiles of galaxies with and
without tidal features in stellar mass bins, this time without star subtraction. While performing star subtraction can alter the measured radii and colours of galaxies, the overall shapes of the colour profiles and the variations in colour between galaxies with and without tidal features remain unchanged.

\bsp	
\label{lastpage}
\end{document}